\documentclass[aps, prd, twocolumn, nofootinbib, preprintnumbers, showpacs]{revtex4}
\usepackage{graphicx}
\usepackage{amsmath}
\usepackage{multirow}
\usepackage{bm}
\usepackage{color}
\def\fsl#1{\setbox0=\hbox{$#1$}                 
   \dimen0=\wd0                                 
   \setbox1=\hbox{/} \dimen1=\wd1               
   \ifdim\dimen0>\dimen1                        
      \rlap{\hbox to \dimen0{\hfil/\hfil}}      
      #1                                        
   \else                                        
      \rlap{\hbox to \dimen1{\hfil$#1$\hfil}}   
      /                                         
   \fi}                                         %

\begin{document}
\title{Constraints on Mass Spectrum of Fourth Generation Fermions 
and Higgs Bosons}
\author{Michio Hashimoto}
 \email{michioh@post.kek.jp}
  \affiliation{
   Theory Center, Institute of Particle and Nuclear Studies,\\
   High Energy Accelerator Research Organization (KEK), \\
   1-1 Oho, Tsukuba, Ibaraki 305-0801, Japan}
\date{\today}
\preprint{KEK-TH-1348}

\begin{abstract}
We reanalyze constraints on the mass spectrum of 
the chiral fourth generation fermions and the Higgs bosons
for the standard model (SM4) and the two Higgs doublet model (THDM).
We find that the Higgs mass in the SM4 should be larger than 
roughly the fourth generation up-type quark mass, while 
the light CP even Higgs mass in the THDM can be smaller. 
Various mass spectra of the fourth generation fermions and 
the Higgs bosons are allowed.
The phenomenology of the fourth generation models is still rich.
\end{abstract}

\maketitle

\section{Introduction}

Repetition of the generation structure of quarks and leptons
is a great mystery in particle physics.
Although three generation models are widely accepted, 
the basic principle of the standard model (SM) allows 
the sequential fourth generation (family)~\cite{Frampton:1999xi,Holdom:2009rf}.
Also, the electroweak precision data does not exclude completely 
existence of the fourth family~\cite{He:2001tp,Novikov:2002tk,Kribs:2007nz}.
Since the LHC has a discovery potential for the fourth generation quarks
at early stage~\cite{AguilarSaavedra:2005pv},
we may explore this possibility more seriously.

If the fourth generation exists,
it is well-known that the condensate of the fourth generation quarks
$t'$ and $b'$ can dynamically trigger the electroweak symmetry breaking 
(EWSB)~\cite{4family}.
In such a scenario, multiple composite Higgs bosons can naturally 
emerge as the scalar bound states of $t'$, $b'$ and 
other heavy fermions such as the top quark $t$ and the fourth family 
leptons $\tau'$ and $\nu'$~\cite{Hashimoto:2009xi,Hashimoto:2009ty}.
When the composite Higgs bosons composed of $t$, $\tau'$ and $\nu'$
are too heavy and hence inaccessible at Tevatron and LHC, 
the effective theory at the TeV scale will be
a two Higgs doublet model (THDM)~\cite{Higgs-hunter}.
Furthermore, if the extra Higgs bosons
other than the SM-like Higgs are decoupled~\cite{Gunion:2002zf}, 
the effective theory of the THDM is reduced into 
the SM with the fourth family (SM4).

In this paper, we study the SM4 and also 
a THDM with the fourth generation.
We assume Dirac-type neutrinos. 
Models with Majorana mass terms will be studied elsewhere.

The yukawa couplings of the fourth generation have 
the Landau pole, so that the SM4 or the THDM are
applicable up to at most several tens TeV.
In this sense, it is natural to expect the existence of 
some strong dynamics such as topcolor models~\cite{Hill:2002ap} 
behind the SM4/THDM.
Nevertheless, we will not impose the compositeness 
condition~\cite{Bardeen:1989ds,Luty:1990bg},
because we are interested in a wider class of models rather than 
the Nambu-Jona-Lasino type one.

We reanalyze the stability condition(s) of the Higgs potential
for the SM4 and the THDM~\cite{Deshpande:1977rw,Sher:1988mj},
and also impose the tree level unitarity
bounds on the yukawa~\cite{Chanowitz:1978uj} and Higgs quartic 
couplings~\cite{Lee:1977yc,Kanemura:1993hm,Akeroyd:2000wc}.
We then find the cutoff $\Lambda$ at which some new physics 
enters to evade the instability of the Higgs potential or
the perturbative description breaks down owing to appearance of
some strong dynamics.
The cutoff $\Lambda$ should not be so small.
Otherwise, the models are not self-contained at the TeV scale.
Besides the theoretical restriction,
we take into account the constraints on
the oblique parameters~\cite{Peskin:1990zt}.

By varying all masses of the fourth generation fermions and 
the Higgs boson(s) within a reasonable parameter space,
we obtain a set of favorable mass spectra.
Strong correlations among the masses of the fermions
and the Higgs bosons are found.
It turns out that the Higgs mass in the SM4 should be 
larger than roughly the $t'$ mass,
while the light CP even Higgs mass in the THDM 
can be smaller because the dynamics of the extra Higgs quartic couplings
can stabilize the Higgs potential against the negative contributions of
the yukawa couplings.
Another noticeable consequence is 
that the decay channel $\tau' \to \nu' + W^-$ is allowed
in a wide parameter space in both of the SM4 and the THDM.
The decay channel $t' \to b' + W^{(*)}$ is not necessarily excluded.
As for the Higgs, 
a decay channel into a pair of the fourth
generation neutrinos is kinematically open 
in a certain parameter region.
Depending on such possibilities, 
more comprehensive studies should be required.

The paper is organized as follows:
In Sec.~\ref{sec-sm4}, we analyze the SM4.
In Sec.~\ref{sec-2HDM}, the THDM is studied. 
Sec.~\ref{summary} is devoted for summary and discussions.
We show the renormalization group equations (RGE's) for the SM4 and 
the THDM in Appendix~\ref{RGE-sm4} and \ref{sec-RGE-2HDM}, respectively.

\section{SM4}
\label{sec-sm4}

Let us study the SM4,
\begin{equation}
  {\cal L}_{\rm SM4} = {\cal L}_{\rm kin} - {\cal L}_Y
  - m_\phi^2 |\phi|^2 - \lambda |\phi|^4,
\end{equation}
with
\begin{eqnarray}
  {\cal L}_y &=& 
   y_{t} \bar{q}_L^{(3)} t_R \tilde{\phi}
 + y_{b} \bar{q}_L^{(3)} b_R \phi
 + y_{t'} \bar{q}_L^{(4)} t'_R \tilde{\phi}
 + y_{b'} \bar{q}_L^{(4)} b'_R \phi  \nonumber \\
&&
 + y_{\tau} \bar{\ell}_L^{(3)} \tau_R \phi  
 + y_{\nu'} \bar{\ell}_L^{(4)} \nu'_R \tilde{\phi}
 + y_{\tau'} \bar{\ell}_L^{(4)} \tau'_R \phi , 
\end{eqnarray}
where $\phi$ represents the Higgs doublet field, 
$\tilde{\phi}$ is defined by $\tilde{\phi} \equiv i\tau_2\phi^*$,
and $q^{(i)}$ and $\ell^{(i)}$ denote the $i$-th family doublet of 
quarks and leptons, respectively.
We take into account the yukawa couplings of the third and fourth 
generations, and ignore other yukawa couplings 
as well as the neutrino masses other than $\nu'$.
As explicitly shown in ${\cal L}_y$, 
we simply assumed the Dirac-type neutrinos. 

The RGE's for the yukawa and Higgs quartic couplings are 
well-known~\cite{Machacek:1983tz,Hill:1980sq}.
We show a set of the RGE's for the gauge, 
yukawa and Higgs-quartic couplings 
at the one-loop approximation in Appendix~\ref{RGE-sm4}.

We explore the cutoff scale $\Lambda$ of the SM4 at which 
some new physics or 
nonperturbative dynamics emerges.
The point is that the yukawa coupling has the Landau pole
at a certain energy scale $\Lambda_y$ and 
only an intermediate mass range of the Higgs boson 
is allowed by the triviality and instability bounds.
Before the full one-loop calculation,
we schematically describe nature of the RGE's.

Let us solve analytically the RGE's 
under the following crude approximation.

The electroweak gauge couplings are negligible.
Although the QCD coupling is not so small, 
it behaves like a constant in the energy scale 
${\cal O}(\mbox{1-10})$ TeV.
On the other hand, the yukawa couplings for the fourth generation 
run very quickly and diverge at the Landau pole.
Thus we may ignore all of the gauge couplings at the zeroth approximation.
For simplicity, we may neglect $y_t$ and 
also assume that all of the fourth generation yukawa couplings 
are the same as $y_4$, 
although it is unrealistic because owing to a relatively heavy Higgs,
the $T$-parameter constraint requires appropriate mass differences of 
the fourth generation fermions, which will be taken into account 
in the full analysis of the one-loop RGE's.

Under the above crude approximation,
the RGE for $y_4$ is given by
\begin{equation}
  (16 \pi^2) \mu \frac{\partial }{\partial \mu} y_4 = 8 y_4^3,
\end{equation}
and the solution is immediately found as
\begin{equation}
 \frac{1}{y_4^2(\mu)}-\frac{1}{y_4^2(\mu_0)}=-\frac{1}{\pi^2}\ln\mu/\mu_0,
 \label{sol-y4}
\end{equation}
where $\mu_0$ is an arbitrary scale.
The universal fermion mass $m_4$ is defined by
$m_4=y_4(\mu=m_4)v/\sqrt{2}$. 
By definition of the Landau pole $\Lambda_y$, 
$1/y_4^2(\mu=\Lambda_y) = 0$ and then
we obtain the relation between $\Lambda_y$ and $m_4$ as
\begin{equation}
  \Lambda_y = m_4 e^{\frac{v^2 \pi^2}{2 m_4^2}},
\end{equation}
where $v\;(=246\mbox{ GeV})$ is the vacuum expectation value (VEV)
of the Higgs.
Numerically, it yields
\begin{equation}
  \Lambda_y=8\,(10)\;\mbox{TeV}, \; 3\,(3)\;\mbox{TeV}, 
  \; 2\,(2)\;\mbox{TeV}, 
\end{equation}
for
\begin{equation}
 m_4=300\;\mbox{GeV}, \; 400\;\mbox{GeV}, \; 500\;\mbox{GeV} \, . 
\end{equation}
Compared with the full one-loop calculation
(the values in the parentheses),
the approximation works in fact.

\begin{figure*}[t]
  \begin{center}
  \resizebox{0.4\textheight}{!}{\includegraphics{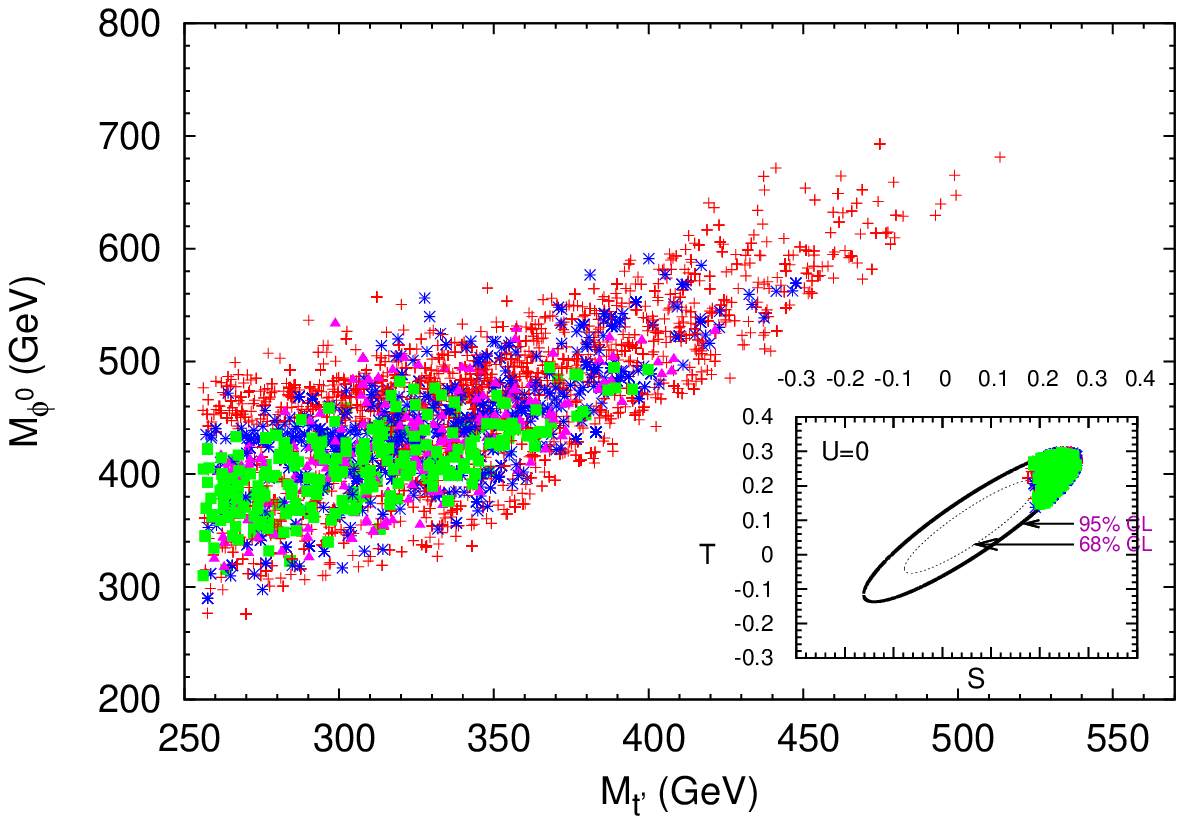}}
  \quad
  \resizebox{0.4\textheight}{!}{\includegraphics{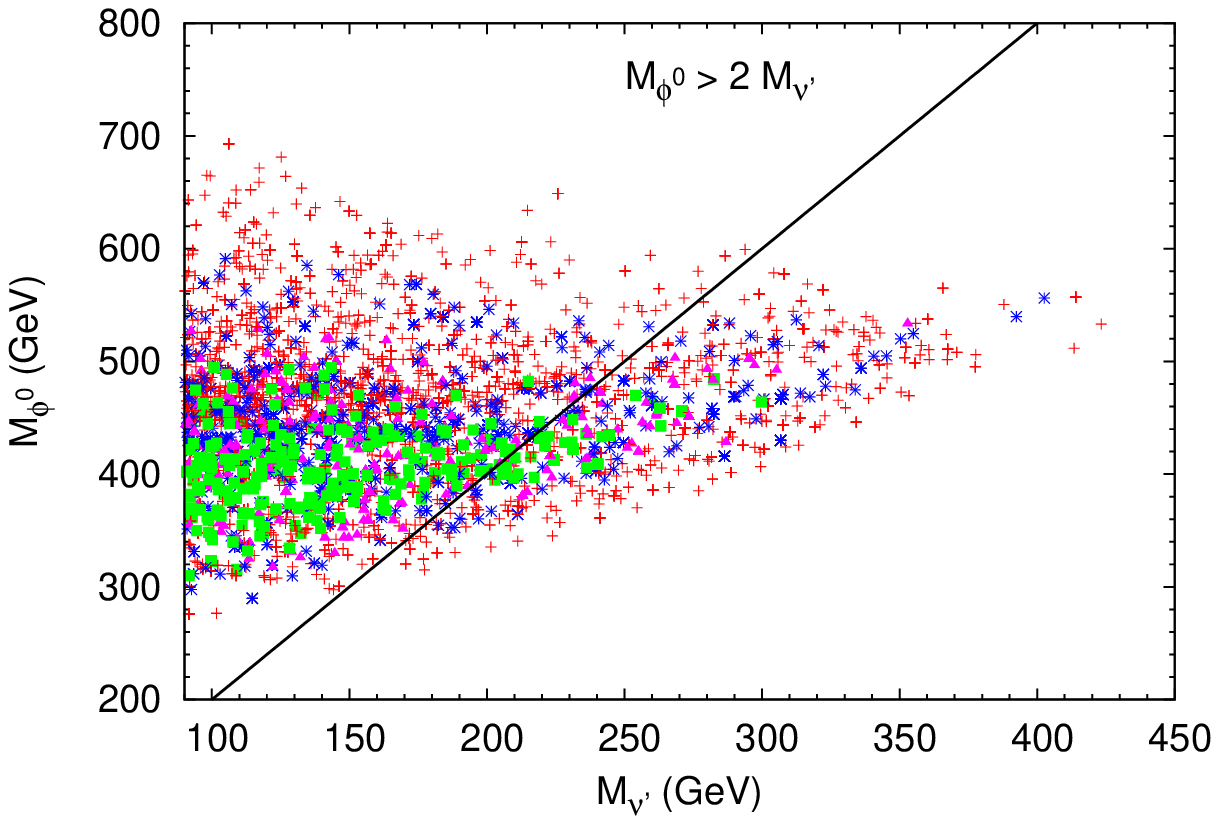}}
  \end{center}
  \caption{Scatter plots of $M_{t'}$ v.s. $M_{\phi^0}$ (left) and
  $M_{\nu'}$ v.s. $M_{\phi^0}$ (right).
  The data points are the same in both figures.
  We varied 
  $256 \mbox{ GeV} < M_{t'} < 552 \mbox{ GeV}$,
  $255 \mbox{ GeV} < M_{b'} < 552 \mbox{ GeV}$,
  $100.8 \mbox{ GeV} < M_{\tau'} < 1.23 \mbox{ TeV}$, 
  $90.3 \mbox{ GeV} < M_{\nu'} < 1.23 \mbox{ TeV}$, and
  $114 \mbox{ GeV} < M_{\phi^0} < 873 \mbox{ GeV}$, 
  without any prejudice. 
  We took into account all of the 40 patterns of the mass spectrum of
  the fermions and the corresponding threshold effects.
  The red, blue, magenta and green points correspond to 
  the cutoff $\Lambda$, 
  $2\mbox{ TeV} \leq \Lambda < 3\mbox{ TeV}$, 
  $3\mbox{ TeV} \leq \Lambda < 4\mbox{ TeV}$, 
  $4\mbox{ TeV} \leq \Lambda < 5\mbox{ TeV}$ and
  $\Lambda \geq 5\mbox{ TeV}$, respectively.
  Below the cutoff scale $\Lambda$, the Higgs potential is stable
  and the perturbation is applicable.
  The data points are within the 95\% C.L. limit of the $S$ and $T$
  parameters.
  In the inset of the left figure, 
  we showed the $(S,T)$-contour and the data points
  which we used for the scatter plot.
  \label{mtp-mh}}
\end{figure*}

As for the Higgs sector, within the above crude approximation,
the RGE for $\lambda$ is
\begin{equation}
  (16 \pi^2) \mu \frac{\partial }{\partial \mu} \lambda =
   24 \lambda^2 + \bigg(\,32 y_4^2 \lambda -16 y_4^4\,\bigg)\theta(\mu-m_4),
\end{equation}
where we explicitly treated the fermion contributions
to the $\beta$-function in the $\theta$-function,
i.e., below the threshold of $m_4$,
the theory is matched to the SM without the fourth generation. 
We can easily find that the following quantity is the RGE invariant:
\begin{equation}
 \eta \equiv
 \frac{\tilde{\lambda}(\mu)-\xi_+}{\tilde{\lambda}(\mu)-\xi_-}
  (y_4(\mu))^{-2\sqrt{7}},
 \quad (\mu > \max(m_4,m_{\phi^0})),
\end{equation}
where $m_{\phi^0} (=\sqrt{2\lambda(\mu=m_{\phi^0})}v)$ 
is the mass of the physical Higgs boson $\phi^0$ 
and also
\begin{equation}
  \tilde{\lambda}(\mu) \equiv \frac{\lambda(\mu)}{y_4^2(\mu)}, \quad
  \xi_\pm \equiv \frac{-1\pm\sqrt{7}}{3} \, .
\end{equation}
Note that when $\eta > 0$, $\lambda$ goes to infinity at 
the scale $\Lambda_\lambda [=\Lambda_y \exp(-\pi^2 \eta^{1/\sqrt{7}})]$,
while it does to zero at the scale 
$\Lambda_{\rm inst}[=\Lambda_y \exp(-\pi^2 \{\xi_- \eta/\xi_+\}^{1/\sqrt{7}})]$,
when $\eta < 0$.

For $m_{\phi^0} < m_4$, the RGE for $\lambda$ develops 
only by the $\lambda^2$-term in the region $m_{\phi^0} < \mu < m_4$,
so that it does not encounter instability in this region.
For $m_{\phi^0} > m_4$, we do not need to care about 
the above threshold effects.
Then, in terms of $m_4$ and $m_{\phi^0}$,
the scale $\Lambda_{{\rm inst}}$ at which 
$\lambda(\mu = \Lambda_{{\rm inst}}) = 0$ 
is given by
\begin{equation}
  \Lambda_{{\rm inst}} = m_4\exp\bigg[\,
  \frac{\pi^2 v^2}{2m_4^2}\bigg\{
  1 - \left(\frac{1-\zeta_1\frac{m_{\phi^0}^2}{8m_4^2}}
                 {1+\zeta_2\frac{m_{\phi^0}^2}{8m_4^2}}\right)^{\frac{1}{\sqrt{7}}}
  \bigg\}\,\bigg] , 
\end{equation}
with
\begin{eqnarray}
  \zeta_1 &\equiv& 
  \frac{\sqrt{7}+1}
       {1+\frac{3m_{\phi^0}^2}{4\pi^2 v^2}\ln\frac{m_{\phi^0}}{m_4}},\\
  \zeta_2 &\equiv& 
  \frac{\sqrt{7}-1}
       {1+\frac{3m_{\phi^0}^2}{4\pi^2 v^2}\ln\frac{m_{\phi^0}}{m_4}} , 
\end{eqnarray}
for $m_{\phi^0} < m_4$, and
\begin{equation}
  \Lambda_{{\rm inst}} = m_4\exp\bigg[\,
  \frac{\pi^2 v^2}{2m_4^2} - t_{\phi^0}
  \left(\frac{1-\frac{(\sqrt{7}+1)m_{\phi^0}^2}{4\pi^2 v^2}t_{\phi^0}}
             {1+\frac{(\sqrt{7}-1)m_{\phi^0}^2}{4\pi^2 v^2}t_{\phi^0}}
  \right)^{\frac{1}{\sqrt{7}}}\,\bigg] ,
\end{equation}
with
\begin{equation}
  t_{\phi^0} \equiv \ln \frac{\Lambda_y}{m_{\phi^0}}  
  = \ln\frac{m_4}{m_{\phi^0}} + \frac{v^2 \pi^2}{2m_4^2} ,  
\end{equation}
for $m_{\phi^0} > m_4$.
Similarly, the Landau pole $\Lambda_\lambda$
for $\lambda$, i.e., $\lambda(\mu=\Lambda_\lambda)=\infty$, 
is given by
\begin{equation}
  \Lambda_\lambda = m_{\phi^0}\exp\bigg[\,
   t_{\phi^0} \left\{1 - \left(1 - \zeta_\lambda \right)^{\frac{1}{\sqrt{7}}}
   \right\}\,\bigg],
\end{equation}
with
\begin{equation}
  \zeta_\lambda \equiv
  \frac{2\sqrt{7}}{\frac{3m_{\phi^0}^2 t_{\phi^0}}
                        {2\pi^2 v^2}+\sqrt{7}+1} , 
\end{equation}
for $m_{\phi^0} > m_4$. 
We find that the solution $\Lambda_\lambda$ for $m_{\phi^0} < m_4$
is phenomenologically unacceptable.

\begin{figure*}[t]
  \begin{center}
  \resizebox{0.4\textheight}{!}{\includegraphics{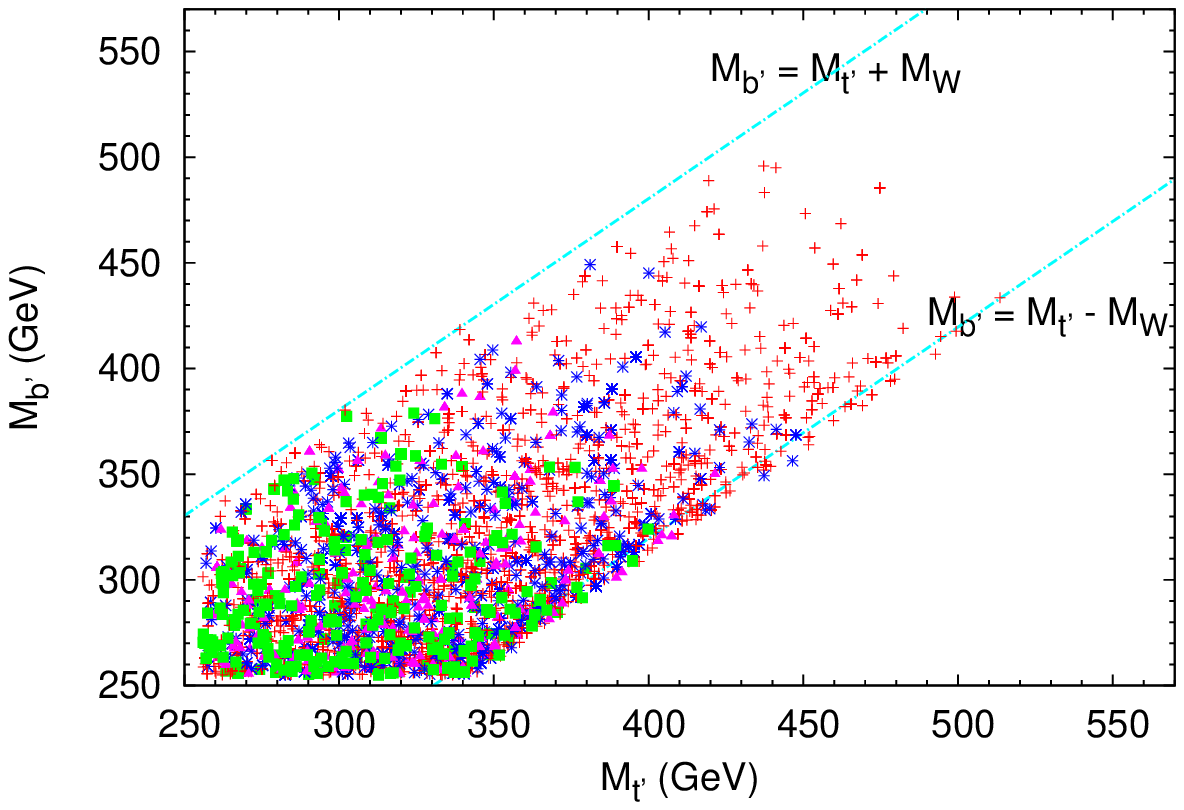}}
  \quad
  \resizebox{0.4\textheight}{!}{\includegraphics{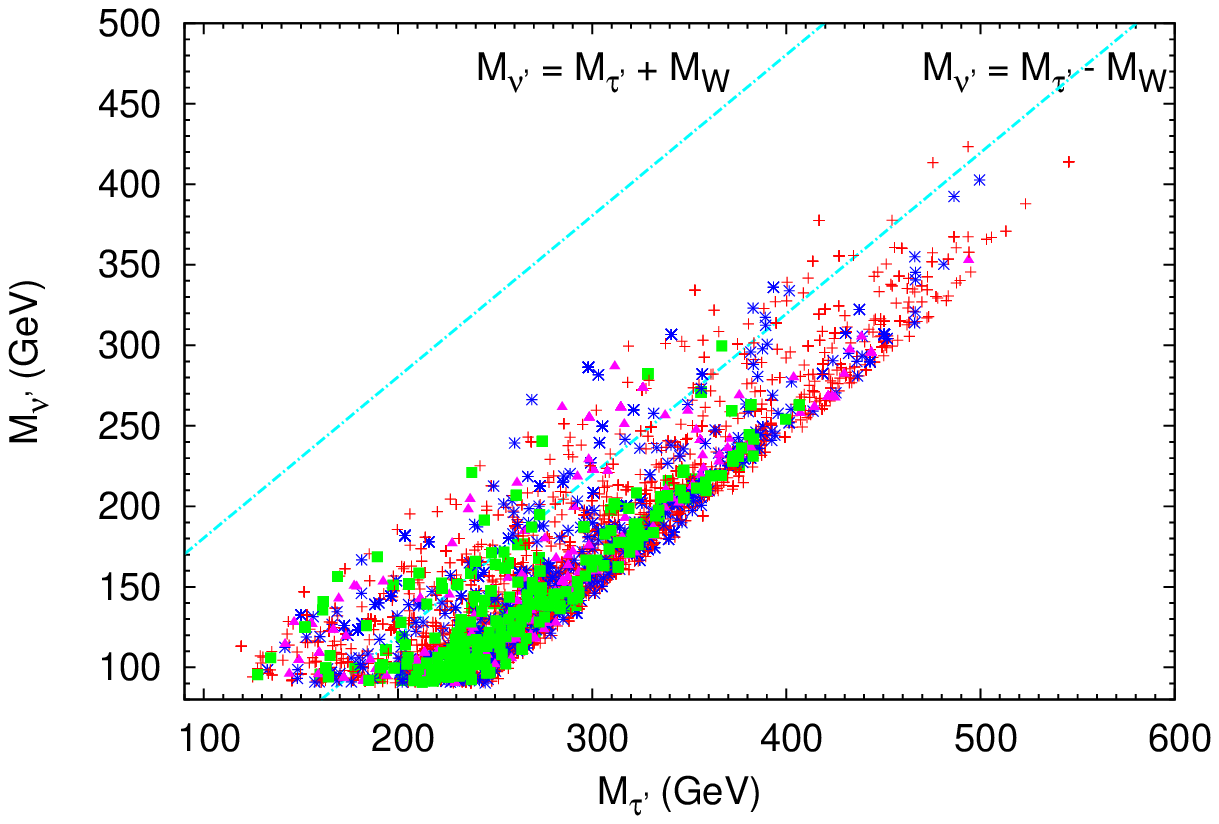}} \\
  \resizebox{0.4\textheight}{!}{\includegraphics{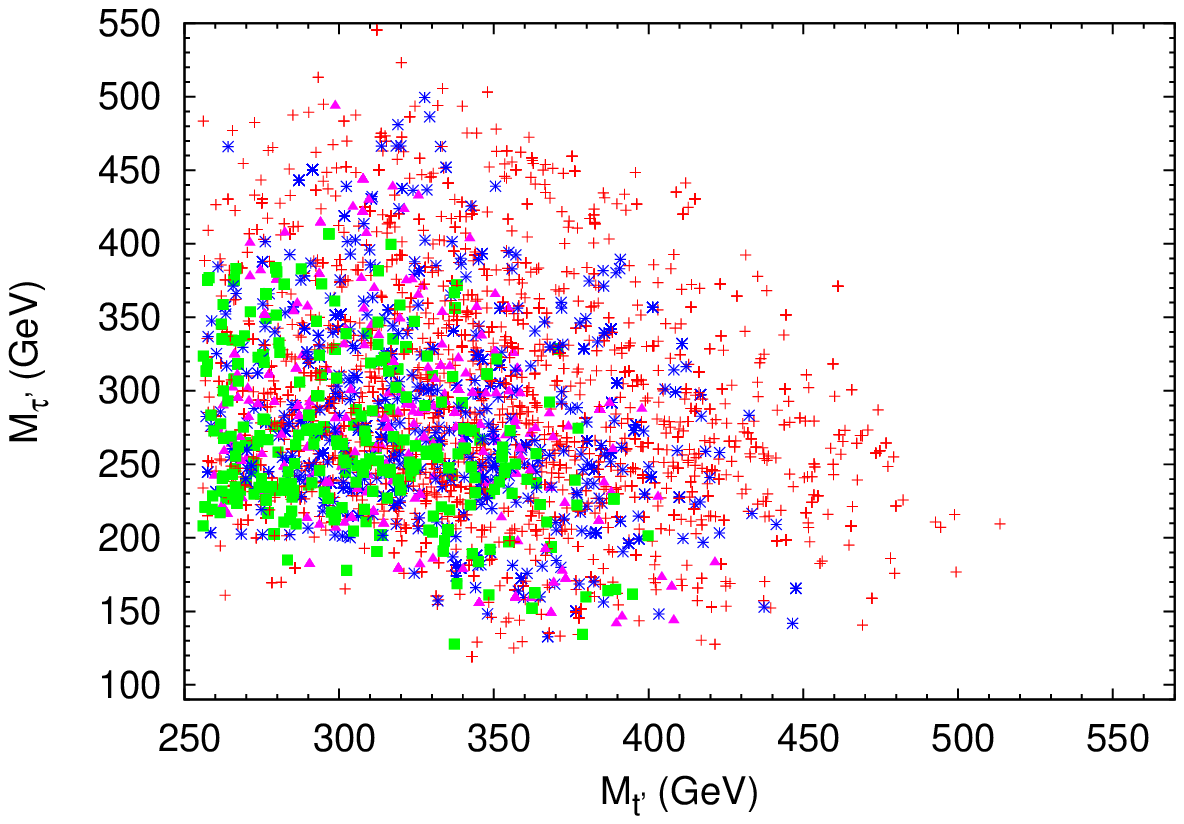}}
  \quad
  \resizebox{0.4\textheight}{!}{\includegraphics{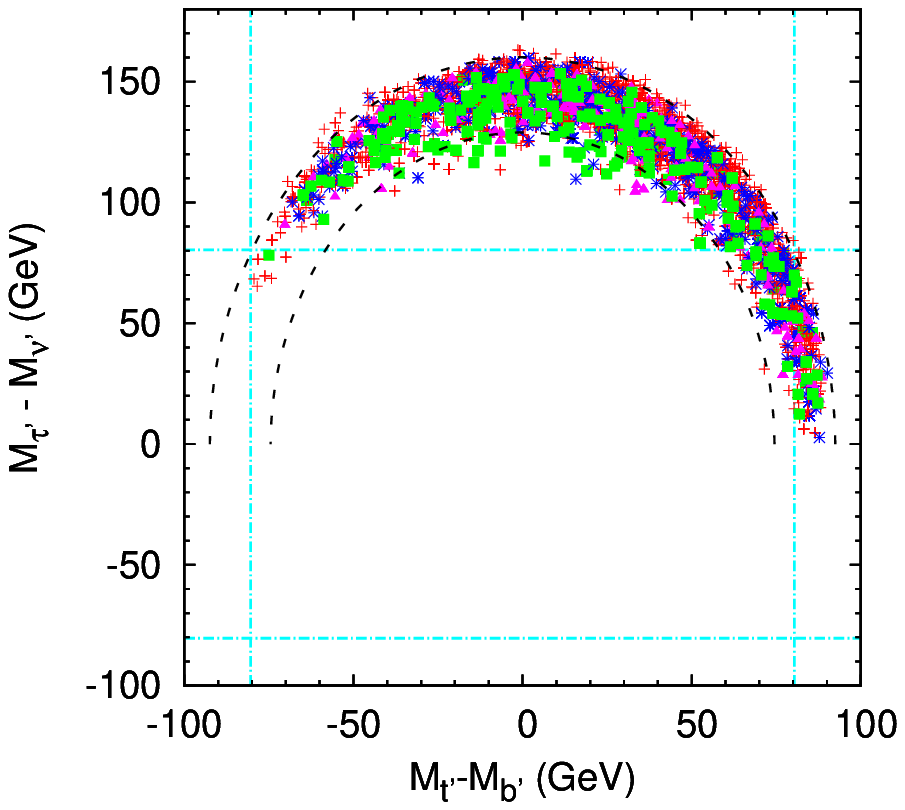}}
  \end{center}
  \caption{$M_{t'}$ v.s. $M_{b'}$ (upper left),
  $M_{\tau'}$ v.s. $M_{\nu'}$ (upper right).
  $M_{t'}$ v.s. $M_{\tau'}$ (lower left) and
  $M_{t'}-M_{b'}$ v.s. $M_{\tau'}-M_{\nu'}$ (lower right).
  The data points are the same as those in Fig.~\ref{mtp-mh}.
  In the upper figures, the blue dashed lines correspond to
  $M_{b'(\nu')} = M_{t'(\tau')}\pm M_W$.
  The blue lines in the lower right correspond to $\pm M_W$.
  Notice that the decay channels $t' \to b' + W^+$ and
  $\tau' \to \nu' + W^-$ are kinematically allowed in the parameter regions
  $M_{t'}-M_{b'} > M_W$ and  $M_{\tau'}-M_{\nu'} > M_W$, respectively.
  In the lower right, the dashed semicircles correspond to 
  the relation (\ref{m4diff}) with $M_{\phi^0}=300,500$ GeV.
  \label{mtp-mbp}}
\end{figure*}

Numerically, with fixing $m_4=300\mbox{ GeV}$, we find
\begin{equation}
  \Lambda_{{\rm inst}}= 0.44\, (0.44) \;\mbox{TeV}, 
  \; 0.71\, (0.74)\;\mbox{TeV}, \; 1.6\, (2.1)\;\mbox{TeV}, 
\end{equation}
for 
\begin{equation}
 m_{\phi^0}= 200\;\mbox{GeV}, \; 300\;\mbox{GeV}, \; 400\;\mbox{GeV}, 
\end{equation}
and
\begin{equation}
  \Lambda_\lambda = 4.3\,(3.7)\;\mbox{TeV}, \; 2.5\,(2.4)\;\mbox{TeV}, 
  \; 2.1\,(2.1)\;\mbox{TeV}, 
\end{equation}
for 
\begin{equation}
 m_{\phi^0}=500\;\mbox{GeV}, \; 600\;\mbox{GeV}, \; 700\;\mbox{GeV} ,  
\end{equation}
where the values in the parentheses are the full one-loop results.
The approximation works well.

In passing, the compositeness conditions~\cite{Bardeen:1989ds},
$1/y_4^2 \to 0$ and $\lambda/y_4^4 \to 0$, require $\eta=0$
and then we get the relation between $m_4$ and $m_{\phi^0}$;
\begin{equation}
  \frac{m_{\phi^0}^2}{2\pi^2 v^2} t_{\phi^0}=\xi_+, \quad (m_{\phi^0} > m_4) \, .
\end{equation}
Numerically, it gives
\begin{equation}
  m_{\phi^0}=480\mbox{ GeV}, 583\mbox{ GeV}, 713\mbox{ GeV}, 
\end{equation}
for
\begin{equation}
  m_4=300\mbox{ GeV}, 350\mbox{ GeV}, 400\mbox{ GeV} \, . 
\end{equation}

We now proceed to perform the full analysis of the one-loop RGE's.

For a quark $q$, we read the $\overline{\rm MS}$-mass via~\cite{pdg}
\begin{equation}
  \hat{m}_q (\mu=\hat{m}_q) = 
   M_q \left(1-\frac{4\alpha_s}{3\pi}+{\cal O}(\alpha_s^2)\right),
\end{equation}
where $M_q$ and $\hat{m}_q$ denote the pole and $\overline{\rm MS}$-masses,
respectively.
We used $\alpha_s(M_Z)=0.118$.
For leptons and the Higgs, the tree level formula is utilized.
We vary the fermion masses~\cite{pdg},
$256 \mbox{ GeV} < M_{t'} < \sqrt{8\pi/5}v$, 
$255 \mbox{ GeV} < M_{b'} < \sqrt{8\pi/5}v$, 
$100.8 \mbox{ GeV} < M_{\tau'} < \sqrt{8\pi}v$, and
$90.3 \mbox{ GeV} < M_{\nu'} < \sqrt{8\pi}v$
without any prejudice, 
where $\sqrt{8\pi/5}v \simeq 552\mbox{ GeV}$ and 
$\sqrt{8\pi}v \simeq 1.23\mbox{ TeV}$ are the perturbative
unitarity bounds for quarks and leptons, respectively~\cite{Chanowitz:1978uj}.
We took into account all of the 40 patterns of 
the mass spectrum of the fermions\footnote{
Owing to the mass bounds $M_{t'}, M_{b'} > M_t$,
the possible patterns are reduced into 
40 from $5!=120$.
}
and the corresponding threshold effects.
For the Higgs mass, we survey the parameter space,
$114 \mbox{ GeV} < M_{\phi^0} < \sqrt{4\pi}v \; (\simeq 873\mbox{ GeV})$.
Imposing the perturbative unitarity bounds on all yukawa couplings, 
and the stability and triviality bound on $\lambda$, 
$0 < \lambda(\mu) < 2\pi$,
we can estimate the theoretical cutoff scale $\Lambda$ for the SM4.

We also take into account the constraints from
the oblique parameters~\cite{Peskin:1990zt}.
In order to suppress the $S$-parameter, $M_{t'} > M_{b'}$ and/or
$M_{\tau'} > M_{\nu'}$ are favorable.
Although it increases the $T$-parameter, this is rather nice, 
because a relatively heavy Higgs pulls down $T$~\cite{He:2001tp,Kribs:2007nz}.
As for estimate of $S$ and $T$,
we follow the LEP EWWG~\cite{LEP:2005ema}.
We obtain the central value as $(S,T)=(0.06,0.08)$,
where the SM point is normalized to $(S,T)=(0,0)$ and 
the top mass $M_t = 173.1$~GeV and 
the reference Higgs mass $M_{{\phi^0},ref}=117$ GeV are used.
The relevant experimental values in the estimate are
$M_Z = 91.1875 \pm 0.0021$~GeV, 
$\Delta\alpha_{{\rm had}}^{(5)} = 0.02758 \pm 0.00035$,
$M_W = 80.399 \pm 0.0025$~GeV,
$\sin^2\theta_{{\rm eff}} = 0.23153 \pm 0.00016$, and
$\Gamma_\ell = 83.985 \pm 0.086$~MeV~\cite{Collaboration:2008ub}. 

In Fig.~\ref{mtp-mh}, we depict scatter plots $M_{t'}$ v.s. $M_{\phi^0}$ 
and $M_{\nu'}$ v.s. $M_{\phi^0}$ 
within the 95\% C.L. limit of the $(S,T)$-constraint.
In each point, the fermion masses are different.
We also showed the $(S,T)$-contour and the data points in the inset of
the left of Fig.~\ref{mtp-mh}.
For consistency of the model, 
the cutoff scale should not be so small.
In the figure, we took the cutoff $\Lambda \geq 2$~TeV.

We find that the theoretical lower bound of the Higgs mass is
$M_{\phi^0} \gtrsim M_{t'}$, when $\Lambda \geq 2$~TeV.
(If we take $\Lambda \geq 1$~TeV, slightly lower values of $M_{\phi^0}$
are allowed, $M_{\phi^0} \gtrsim M_{t'} - \mbox{50 GeV}$.)
Note that the Higgs production via the gluon fusion process is 
considerably enhanced owing to the loop effects of $t'$ and $b'$.
For example, the cross section $\sigma_{gg \to \phi^0}$ 
for $M_{t'}=M_{b'}=0.4$~TeV and $M_{\phi^0}=0.5$~TeV increases 
by a factor of 5. 
Depending on the masses of $t',b'$ and $\phi^0$, 
the enhancement factor varies from 5 to 9.
Consequently, a wider mass range of $M_{\phi^0}$ should be excluded 
at Tevatron.
This potentially excluded mass range is fairly lower than 
the above Higgs mass bound, however~\cite{Schmidt:2009kk}.

In addition, the right of Fig.~\ref{mtp-mh} clearly shows that
the decay channel $\phi^0 \to \bar{\nu}' \nu'$ is opened
in a favorable parameter space.
The importance of this process has been emphasized 
in Ref.~\cite{Kribs:2007nz}, i.e., 
the new signal via $\phi^0 \to \bar{\nu}' \nu' \to 4\ell+E\!\!\!/$,
where $E\!\!\!/$ is the missing energy, can be comparable to 
the rate for $\phi^0 \to ZZ \to 4\ell$.
Moreover, we find that there is a parameter region where
$\phi^0 \to \bar{\tau}' \tau'$ is also kinematically allowed.
In fact, several scenarios are possible.
We show data samples in Table~\ref{tab-1}.

The constraints from the oblique parameters cause strong correlations
between $M_{t'}$ and $M_{b'}$ and also between $M_{\tau'}$ and $M_{\nu'}$,
as shown in the upper left and right of Fig.~\ref{mtp-mbp}. 
The lower left of Fig.~\ref{mtp-mbp} suggests that 
there is no correlation between $M_{t'}$ and $M_{\tau'}$,
as expected.
On the other hand, the fermion mass differences are strongly correlated,
as shown in the lower right of Fig.~\ref{mtp-mbp}. 
This essentially corresponds to the constraint of the $T$-parameter,
\begin{eqnarray}
\lefteqn{\hspace*{-1.5cm}
   \frac{3(M_{t'} - M_{b'})^2}{M_W^2}
 + \frac{(M_{\tau'} - M_{\nu'})^2}{M_W^2}
} \nonumber \\
&&
 \approx \mbox{(1.3--2.0)} + 1.4 \ln \frac{M_{\phi^0}}{M_{\phi^0,ref}} \,.
 \label{m4diff}
\end{eqnarray}
We depicted it with 
$M_{\phi^0}=300\mbox{ GeV}, \; 500\mbox{ GeV}$
and $M_{{\phi^0},ref}=117$~GeV
in the semicircles of the lower right of Fig.~\ref{mtp-mbp}. 
The $S$-parameter constraint also suggests that
the parameter region $M_{\tau'} > M_{\nu'}$ and $M_{t'} > M_{b'}$ 
is favorable.
(Within the 95\% C.L. limit, $M_{t'} < M_{b'}$ is also possible
with paying cost of a worse $\chi^2$,
as shown in the lower right of Fig.~\ref{mtp-mbp}. )

We emphasize that in a wide parameter region, 
we find $M_{\tau'} > M_{\nu'} + M_W$, i.e.,
the decay channel $\tau' \to \nu'+W^-$ is allowed.
(The situation is unchanged, even if we take $\Lambda \geq 1$~TeV.)
Also, $t' \to b' + W^{(*)}$ is possible.
These do not necessarily contradict the results
in Ref.~\cite{Kribs:2007nz}:
Since the Higgs is inevitably heavy in our approach,
the $T$-parameter constraint requires a larger mass difference
of the fermions than that of Ref.~\cite{Kribs:2007nz}.
(The $\chi^2$ is a bit worse, however.)

The implications of Figs.~\ref{mtp-mh}--\ref{mtp-mbp} are obvious:
If $t'$ and/or $b'$ are discovered at the Tevatron and/or LHC,
the Higgs mass will be suggested under the assumption of the SM4.
On the other hand, if the LHC excludes $M_{t',b'} \lesssim 500$~GeV,
only few points survive when we take $\Lambda \geq 2$~TeV.
I.e., a model with the cutoff $\Lambda < 2$~TeV
or a nonperturbative regime will be left to be explored.

\begin{table}
  \begin{center}
  \begin{tabular}{c||c|cc|cc|c|cc|} \hline
   & $M_{\phi^0}$ & $M_{t'}$ & $M_{b'}$ & $M_{\nu'}$ & $M_{\tau'}$
   & $\Lambda$ & $S$ & $T$ \\  \hline \hline
  (a1) & 0.47 & 0.36 & 0.34 & 0.092 & 0.23 & 3.5 & 0.20 & 0.22 \\
  (a2) & 0.59 & 0.45 & 0.42 & 0.092 & 0.24 & 2.3 & 0.21 & 0.22 \\ \hline
  (b1) & 0.45 & 0.32 & 0.36 & 0.098 & 0.23 & 8.0 & 0.23 & 0.27 \\
  (b2) & 0.55 & 0.40 & 0.41 & 0.13 & 0.28 & 3.4 & 0.24 & 0.22 \\ \hline
  (c1) & 0.51 & 0.44 & 0.36 & 0.10 & 0.17 & 2.1 & 0.24 & 0.28 \\
  (c2) & 0.65 & 0.50 & 0.42 & 0.097 & 0.18 & 2.1 & 0.25 & 0.29 \\ \hline
  (d1) & 0.39 & 0.32 & 0.34 & 0.093 & 0.23 & 2.4 & 0.21 & 0.23 \\
  (d2) & 0.55 & 0.43 & 0.41 & 0.13 & 0.28 & 2.1 & 0.24 & 0.23 \\ \hline
  (e1) & 0.49 & 0.36 & 0.34 & 0.26 & 0.40 & 2.1 & 0.26 & 0.24 \\ \hline
  \end{tabular}
  \caption{Data samples for several scenarios. The mass unit is TeV.
    For (a1), (b1), (c1), (d1) and (e1), we took the mass bounds 
    $M_{t'} > 311$~GeV and $M_{b'} > 338$~GeV~\cite{CDF-bound},
    whereas we did $M_{t',b'} > 400$~GeV for (a2), (b2), (c2) and (d2).
    For all samples, we took $\Lambda \geq 2$~TeV.
    The criterion for (a1) and (a2) is the $\chi^2$-minimum.
    Similarly, (b1) and (b2) have the largest $\Lambda$
    within the 95\% C.L. limit of the $(S,T)$-constraints.
    For (c1) and (c2), $t' \to b' + W^+$ is possible.
    The samples (d1) and (d2) are most favorable data 
    for $M_{\phi^0} < 2M_{\tau'}$,
    while (e1) is for $M_{\phi^0} < 2M_{\nu'}$.
    We do not have the data sample with $M_{t',b'} > 400$~GeV 
    and $M_{\phi^0} < 2M_{\nu'}$. \label{tab-1}}
  \end{center}
\end{table}

\section{Two Higgs doublet model}
\label{sec-2HDM}

\subsection{Model}

Let us consider the THDM with the fourth generation:
\begin{equation}
  {\cal L}_{\rm THDM} = {\cal L}_{\rm kin} - {\cal L}_Y - V,
\end{equation}
where ${\cal L}_{\rm kin}$ represents the kinetic terms 
of the fermions, the Higgs fields and the SM gauge fields, 
${\cal L}_Y$ denotes the yukawa couplings between fermions and 
the Higgs fields, and $V$ is the Higgs potential.
The yukawa sector, in particular the neutrino one,
is model-dependent.
For simplicity, we assume that the neutrinos have the Dirac masses. 
A model with Majorana neutrinos should be considered separately.
This is, however, out of scope in this paper.

We define the THDM of the type~II (THDM~II) 
with the Dirac neutrinos as follows:
One Higgs doublet ($\Phi_1$) couples to 
the down-type quarks and charged leptons, while
the other ($\Phi_2$) does to the up-type quarks and 
neutral leptons, i.e.,
\begin{eqnarray}
  {\cal L}_Y &=& 
   \sum_{i,j=1}^4 Y_{U}^{ij} \bar{q}_L^{(i)} u_R^{(j)} \tilde{\Phi}_2
 + \sum_{i,j=1}^4 Y_{D}^{ij} \bar{q}_L^{(i)} d_R^{(j)} \Phi_1
   \nonumber \\
&+&
   \sum_{i,j=1}^4 Y_{N}^{ij} \bar{\ell}_L^{(i)} \nu_R^{(j)} \tilde{\Phi}_2
 + \sum_{i,j=1}^4 Y_{E}^{ij} \bar{\ell}_L^{(i)} e_R^{(j)} \Phi_1,
\end{eqnarray}
where $u_R^{(j)}$ represents the right-handed up-type quark
of the $j$-th family and the definitions of $d_R^{(j)}$ etc. are
then self-evident.

The Higgs potential is
\begin{eqnarray}
&& \hspace*{-10mm}
  V =
  m_1^2\Phi_1^\dagger \Phi_1
 +m_2^2\Phi_2^\dagger \Phi_2
 +m_{12}^2 (\Phi_1^\dagger \Phi_2 + \mbox{(h.c.)})
  \nonumber \\ && \hspace*{-7mm}
 +\lambda_1 (\Phi_1^\dagger \Phi_1)^2
 +\lambda_2 (\Phi_2^\dagger \Phi_2)^2
 +\lambda_3 (\Phi_1^\dagger \Phi_1)(\Phi_2^\dagger \Phi_2)
  \nonumber \\ && \hspace*{-7mm}
 +\lambda_4 (\Phi_1^\dagger \Phi_2)(\Phi_2^\dagger \Phi_1)
 +\frac{1}{2}\lambda_5 ((\Phi_1^\dagger \Phi_2)(\Phi_1^\dagger \Phi_2)
 + \mbox{(h.c.)}) ,  
 \label{V_THDM}
\end{eqnarray}
where we do not consider the hard $Z_2$-breaking terms.
Owing to the (softly broken) $Z_2$-symmetry,
the tree-level FCNC is absent~\cite{Glashow:1976nt}.

We do not consider CP violation in the (tree-level) Higgs sector
and hence will take all parameters in the Higgs potential $V$
to be real, so that there are eight parameters.
In the yukawa sector, we assume that the mixing terms between 
the fourth generation and the others are absent, i.e.,
$Y_U^{4k}=Y_U^{k4}=y_{t'}\delta_{4k}$, 
$Y_D^{4k}=Y_D^{k4}=y_{b'}\delta_{4k}$,
$Y_N^{4k}=Y_N^{k4}=y_{\nu'}\delta_{4k}$, and 
$Y_E^{4k}=Y_E^{k4}=y_{\tau'}\delta_{4k}$.
We can then reduce number of parameters.

When the EWSB occurs, three ($G^0$ and $G^\pm$) of 
the eight scalar degrees of freedom are eaten by 
the weak gauge bosons.
The physical mass spectrum then contains
two CP even Higgs bosons $h$ and $H$ defined by $M_h < M_H$,
one CP odd Higgs $A$, and
the charged Higgs pair $H^\pm$, so that 
the original Higgs fields are written in terms of 
the physical degrees of freedom as follows;
\begin{equation}
  \Phi_1 = \frac{1}{\sqrt{2}}
  \left(\begin{array}{c}
   \sqrt{2} (c_\beta G^+ - s_\beta H^+) \\
    c_\beta v  - s_\alpha h + c_\alpha H + i(c_\beta G^0 - s_\beta A)
    \end{array}\right),
\end{equation}
and
\begin{equation}
  \Phi_2 = \frac{1}{\sqrt{2}}
  \left(\begin{array}{c}
   \sqrt{2} (s_\beta G^+ + c_\beta H^+) \\
    s_\beta v  + c_\alpha h + s_\alpha H + i(s_\beta G^0 + c_\beta A)
    \end{array}\right),
\end{equation}
where $\alpha$ is the mixing angle between $h$ and $H$
and the ratio of the VEVs of the two Higgs fields is defined by 
$\tan\beta$.
We also used the notations
$s_\beta \equiv \sin\beta$, $c_\beta \equiv \cos\beta$, and etc..

It is convenient to express 
the quartic couplings $\lambda_{1\mbox{--}5}$ 
through the Higgs masses, the soft $Z_2$-breaking term $m_{12}^2$,
and the mixing angles $\alpha$ and $\beta$: 
\begin{subequations}
\label{lam-mass}
\begin{eqnarray}
 2 \lambda_1 v^2 &=&
  \frac{c_\alpha^2 M_H^2 + s_\alpha^2 M_h^2}{c_\beta^2} - M^2 \tan^2 \beta, \\
 2 \lambda_2 v^2 &=&
  \frac{s_\alpha^2 M_H^2 + c_\alpha^2 M_h^2}{s_\beta^2} - M^2 \tan^{-2} \beta , \\
 \lambda_3 v^2 &=& 
   \frac{s_{2\alpha}}{s_{2\beta}}(M_H^2-M_h^2) + 2M_{H^\pm}^2 - M^2, \\
 \lambda_4 v^2 &=& M_A^2 - 2M_{H^\pm}^2 + M^2, \\
 \lambda_5 v^2 &=& -M^2_{A} +  M^2 , 
\end{eqnarray}
\end{subequations}
where we used $M^2$,
\begin{equation}
  M^2 \equiv \frac{-m_{12}^2}{s_\beta c_\beta} , 
\end{equation}
instead of $m_{12}^2$.
Since we consider a general THDM,
all quartic couplings are independent and
hence free from the MSSM relations~\cite{Higgs-hunter}.

The decoupling limit of the extra Higgs corresponds to 
$M_h^2 \sim {\cal O}(v^2)$ and 
$M_{H,A,H^\pm}^2 \sim M^2 \gg v^2$~\cite{Gunion:2002zf}, 
where $M^2$ is independent of 
the quartic couplings $\lambda_{1\mbox{--}5}$ 
and thus can be taken as some high scale 
without contradict against the perturbative unitarity bound.
In this case, $\sin(\beta-\alpha) \simeq 1$ is also derived.
The low energy effective theory in this limit is reduced into the SM4.

\subsection{Methodology}

\begin{figure*}[t]
  \begin{center}
  \resizebox{0.4\textheight}{!}{\includegraphics{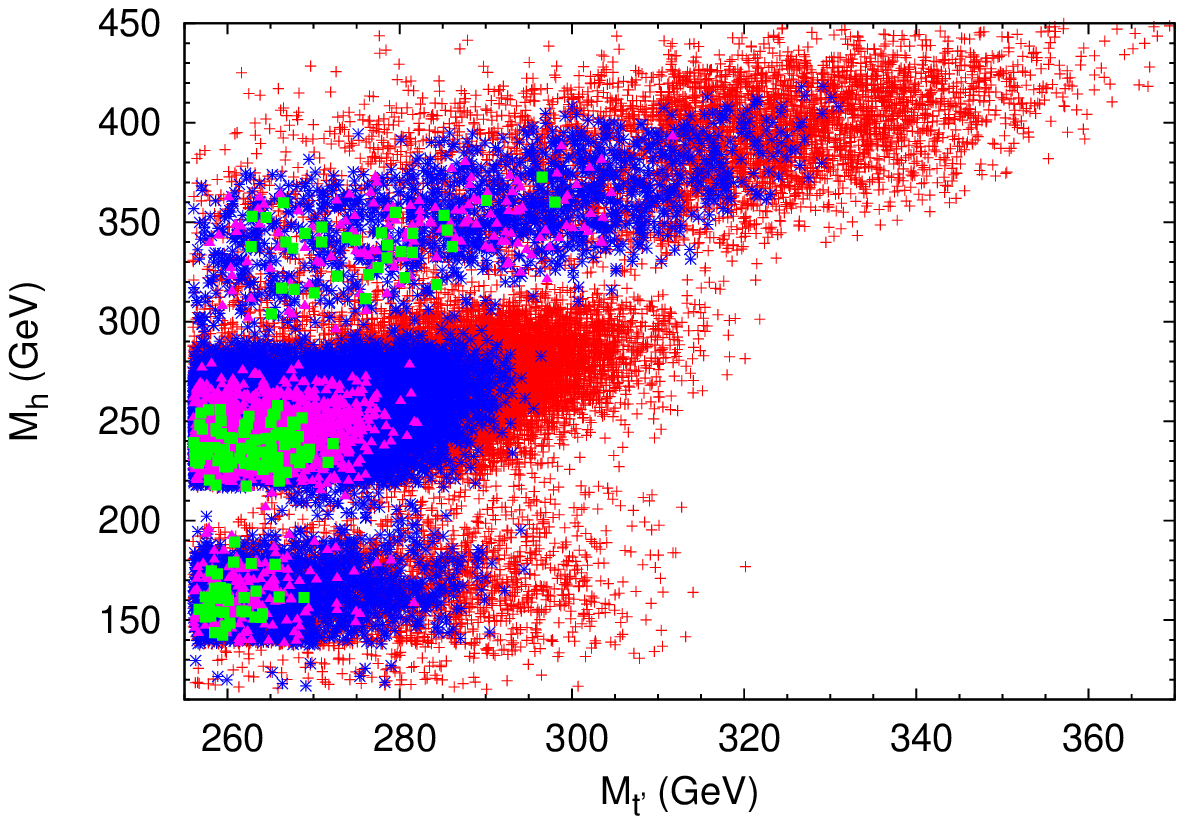}} \quad
  \resizebox{0.4\textheight}{!}{\includegraphics{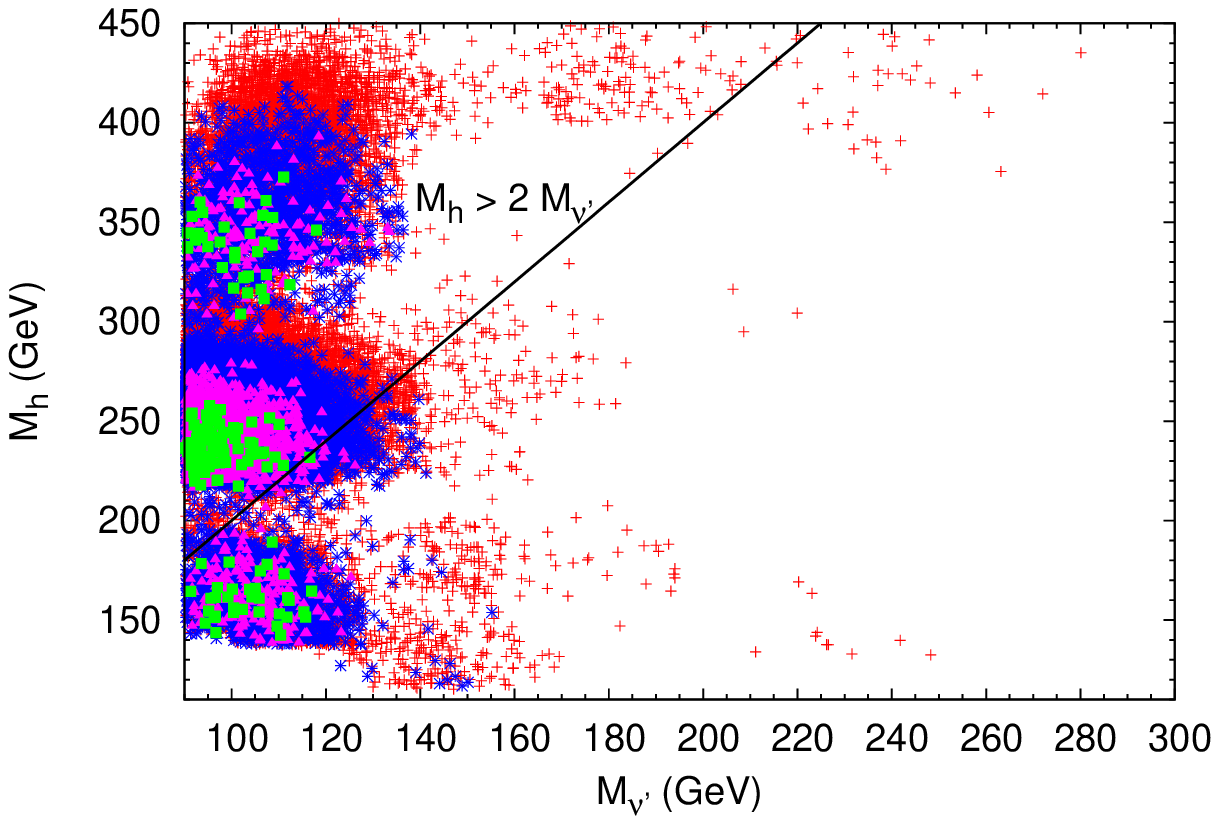}}
  \end{center}
  \caption{Scatter plots of $M_{t'}$ v.s. $M_{h}$ (left)
  and $M_{\nu'}$ v.s. $M_{h}$ (right).
  The data points are the same in both figures.
  We took $\sin(\beta-\alpha)=1$
  and varied 
  $256 \mbox{ GeV} < M_{t'} < 552 \mbox{ GeV}$,
  $255 \mbox{ GeV} < M_{b'} < 552 \mbox{ GeV}$,
  $100.8 \mbox{ GeV} < M_{\tau'} < 1.23 \mbox{ TeV}$, 
  $90.3 \mbox{ GeV} < M_{\nu'} < 1.23 \mbox{ TeV}$,
  $114 \mbox{ GeV} < M_{h} < 1 \mbox{ TeV}$,
  $M_h < M_{H} < 1.5 \mbox{ TeV}$,
  $300 \mbox{ GeV} < M_{H^\pm} < 1 \mbox{ TeV}$,
  $93 \mbox{ GeV} < M_{A} < 1 \mbox{ TeV}$, 
  $0.5 < \tan\beta < 5$, and $|\lambda_5| < \pi$. 
  The red, blue, magenta and green points correspond to 
  the cutoff $\Lambda$, 
  $2\mbox{ TeV} \leq \Lambda < 3\mbox{ TeV}$, 
  $3\mbox{ TeV} \leq \Lambda < 4\mbox{ TeV}$, 
  $4\mbox{ TeV} \leq \Lambda < 5\mbox{ TeV}$ and
  $\Lambda \geq 5\mbox{ TeV}$, respectively.
  All data are within the 95\% C.L. limit of the $S$ and $T$
  parameters.
  \label{mtp-mh-2HDM}}
\end{figure*}

\begin{figure}[t]
  \begin{center}
  \resizebox{0.4\textheight}{!}{\includegraphics{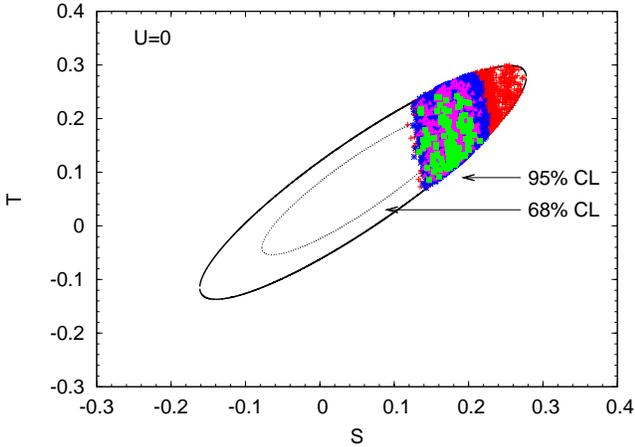}}
  \end{center}
  \caption{The 68\% and 95\% C.L. constraints on the $S$ and $T$ parameters.
  We also showed the data points in Fig.~\ref{mtp-mh-2HDM}.
  \label{S-T-2HDM}}
\end{figure}

Since we assume that all parameters in the Higgs potential $V$
are real, there are eight parameters in the Higgs sector.
A convenient choice is to take
\begin{eqnarray*}
  && M_h, \quad M_H, \quad M_A, \quad M_{H^\pm}, \\
  && \lambda_5, \quad \tan\beta, \quad \sin(\beta-\alpha), \quad 
      v\;(=\mbox{246 GeV}) \, . 
\end{eqnarray*}
In this paper, we fix $\sin(\beta-\alpha)=1$,
at which $h$ is SM-like.
Furthermore, there are four parameters 
corresponding to the pole masses of 
the fourth family quarks and leptons,
\begin{equation*}
  M_{t'}, \quad   M_{b'}, \quad   M_{\nu'}, \quad   M_{\tau'}\,.
\end{equation*}
Basically we search a favorable parameter space
by varying the above ten parameters, 
as in the analysis of the SM4.

One of the problem is the matching condition
between the SM4 and the THDM:
At least in the decoupling limit characterized by
$M_h^2 \sim {\cal O}(v^2)$ and $M_{H,A,H^\pm}^2 \sim M^2 \gg v^2$,
we need to consider the matching of the two theories, SM4 and THDM.
We here note that 
the structure of the yukawa sector as well as the Higgs sector is 
quite different in the two theories.
In the THDM~II, 
the VEV of the Higgs $\Phi_i$ ($i=1,2$) provides the fermion mass $m_i$ 
(for the fourth generation, $m_{1,2}=m_{b',t'}$ or $m_{1,2}=m_{\tau',\nu'}$), 
so that the relation between the mass $m_i$ and 
the yukawa coupling $y_i$ is
\begin{equation}
  m_1 = \frac{y_1 c_\beta}{\sqrt{2}} v , \quad 
  m_2 = \frac{y_2 s_\beta}{\sqrt{2}} v , 
\end{equation}
where $y_{1,2}=y_{b',t'}$ or $y_{1,2}=y_{\tau',\nu'}$ for the fourth generation.
On the other hand, the fermion mass in the SM is given by
\begin{equation}
  m_i = \frac{y_i^{\rm SM}}{\sqrt{2}}\; v , \quad (i=1,2) \, . 
\end{equation}
Besides, the running effects of the yukawa couplings 
are different.
These affect estimate of the cutoff $\Lambda$.

A simple case is the situation $M_H=M_A=M_{H^\pm}$.
By definition, $M_h < M_H$ and thus
we can apply the SM4 up to the scale $\mu=M_H=M_A=M_{H^\pm}$,
where we will identify $h$ to the SM Higgs $\phi^0$.
Above the heavy Higgs scale, $\mu > M_H=M_A=M_{H^\pm}$,
we utilize the THDM description.

For a general mass spectrum of the Higgs bosons,
we handle the problem as follows:

When we randomly generate data of the physical Higgs masses $M_{h,H,A,H^\pm}$,
we define the lightest Higgs mass among them
by $\mu_{LH}^{}$, i.e., $\mu_{LH}^{} \equiv \min (M_h,M_H,M_A,M_{H^\pm})$.
Similarly, the second lightest one is defined by $\mu_{2LH}^{}$.

When $\mu_{LH}^{} = M_h$,
we can regard the theory in the region $\mu_{LH}^{} < \mu < \mu_{2LH}^{}$
as the one Higgs doublet model.
(In $\mu < \mu_{LH}^{}$, the corresponding theory is ``Higgsless''.)
We then improve the yukawa and Higgs quartic couplings 
by using the RGE's of the SM4 up to the scale $\mu_{2LH}^{}$.
In the region $\mu > \mu_{2LH}^{}$, 
the one Higgs description can't be valid.
Thus we employ the matching conditions at $\mu = \mu_{2LH}^{}$, 
\begin{eqnarray}
  y_1^{\rm SM}(\mu=\mu_{2LH}^{}) &=& y_1(\mu=\mu_{2LH}^{})\, c_\beta\, , \\
  y_2^{\rm SM}(\mu=\mu_{2LH}^{}) &=& y_2(\mu=\mu_{2LH}^{})\, s_\beta\, , 
\end{eqnarray}
for the yukawa couplings, and
\begin{eqnarray}
&& \hspace*{-10mm}
   2 \lambda_{\rm SM} (\mu=\mu_{2LH}^{}) = 
   \frac{M^2}{v^2} c_{\beta-\alpha}^2 \nonumber \\
&& 
 + 2 \lambda_1 s_\alpha^2 c_\beta^2 + 2 \lambda_2 c_\alpha^2 s_\beta^2
              - \frac{1}{2}\lambda_{345} s_{2\alpha} s_{2\beta},
  \label{lam-SM-2HDM}
\end{eqnarray}
with $\lambda_{345} \equiv \lambda_3+\lambda_4+\lambda_5$,
for the Higgs quartic couplings,
where $\lambda_{\rm SM}$ represents the SM one.
Eliminating $M_h^2$ from the THDM relation (\ref{lam-mass}) 
and using Eq.~(\ref{lam-SM-2HDM}) instead, 
we obtain the quartic couplings $\lambda_{1\mbox{--}5}$ of the THDM
at the scale $\mu=\mu_{2LH}^{}$.
Practically, we can find $\lambda_{1\mbox{--}5}$
by replacing $M_h^2$ in (\ref{lam-mass}) by the RG improved SM value. 

On the other hand, if $\mu_{LH}^{} \ne M_h$,
the low energy effective theory at the TeV scale is no longer the SM4.
In this case, we may treat the theory as the THDM from the beginning.

In this paper, we do not consider a general case with
$\sin(\beta-\alpha) \ne 1$.
For a full analysis of the THDM,
more sophisticated prescription should be required.

\begin{figure*}[t]
  \begin{center}
  \resizebox{0.4\textheight}{!}{\includegraphics{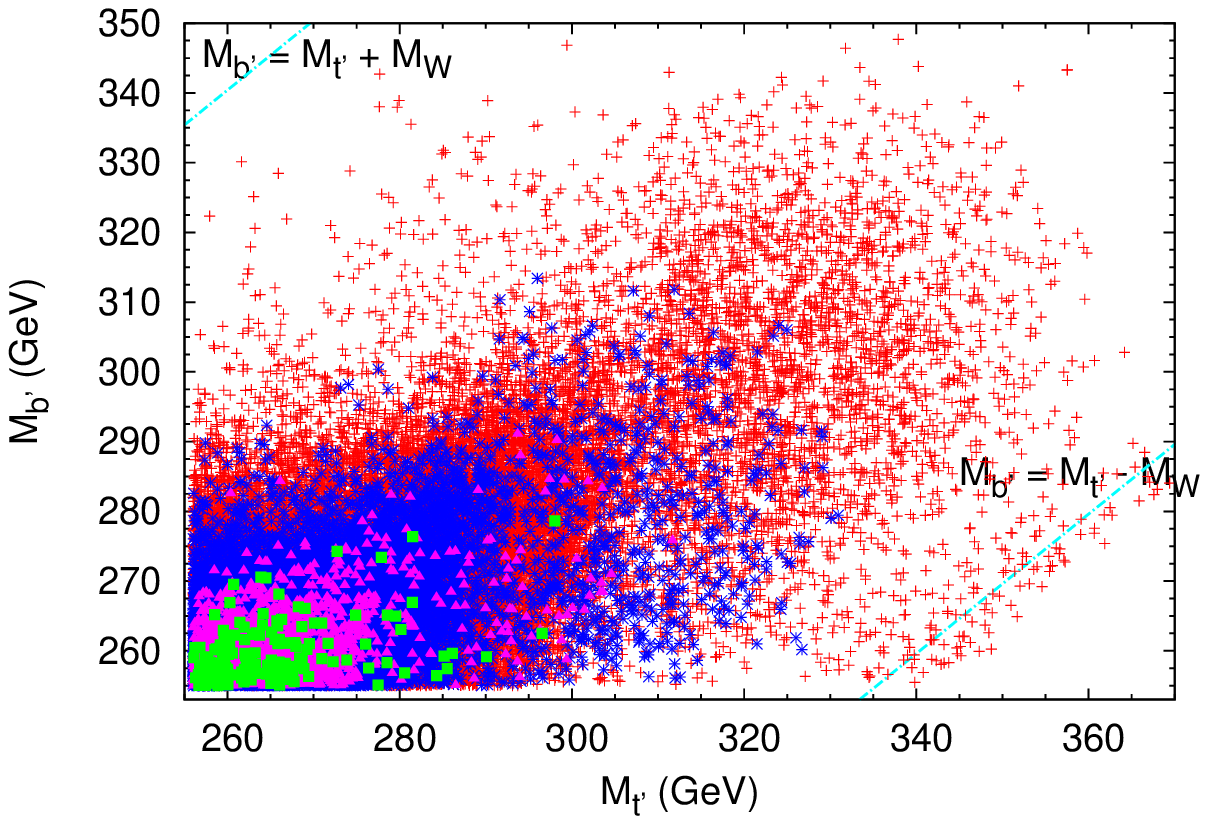}}
  \quad
  \resizebox{0.4\textheight}{!}{\includegraphics{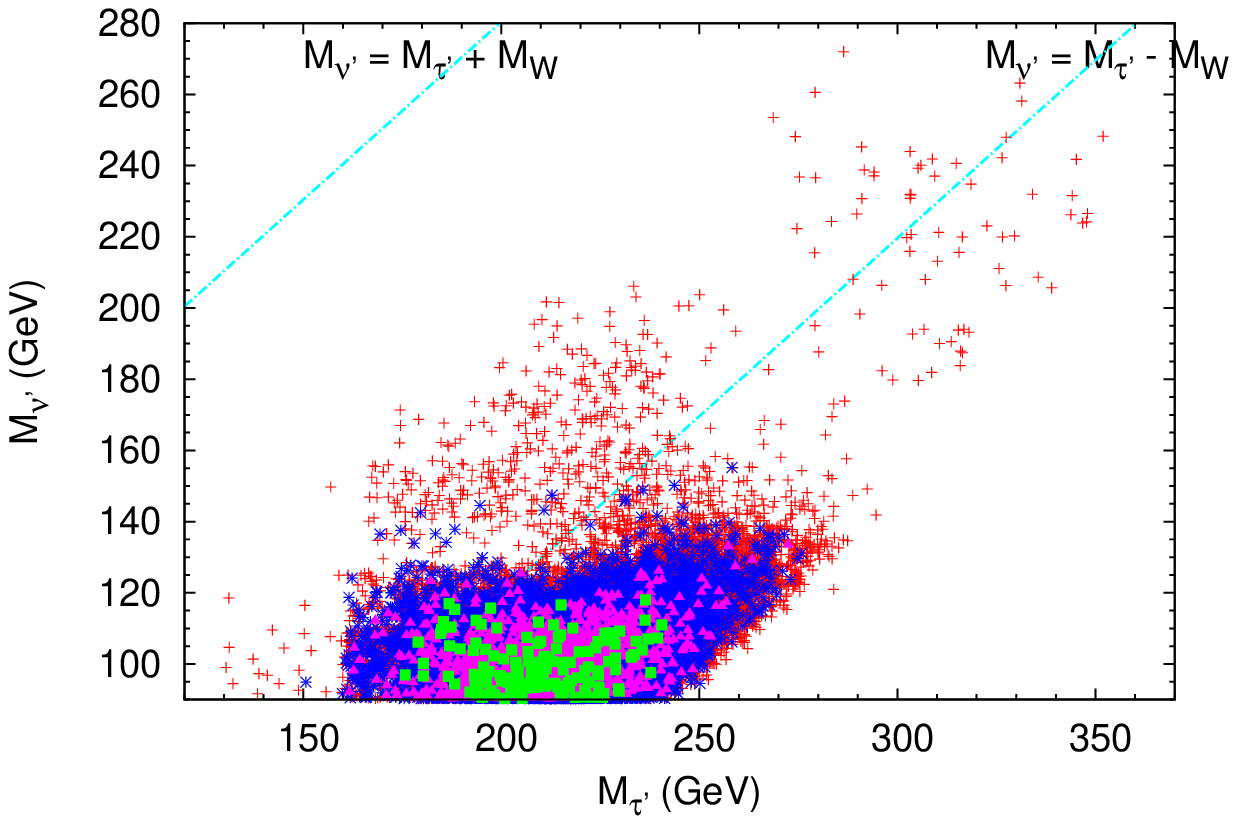}} \\
  \resizebox{0.4\textheight}{!}{\includegraphics{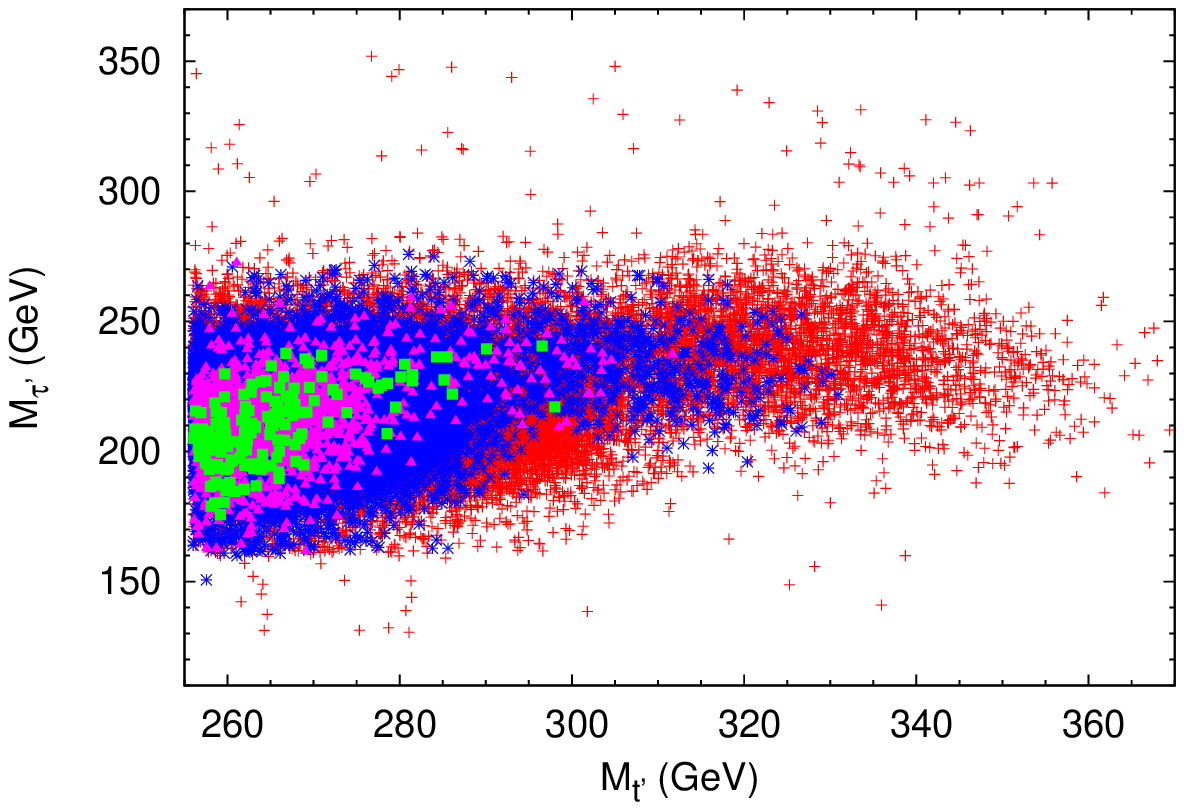}}
  \quad
  \resizebox{0.4\textheight}{!}{\includegraphics{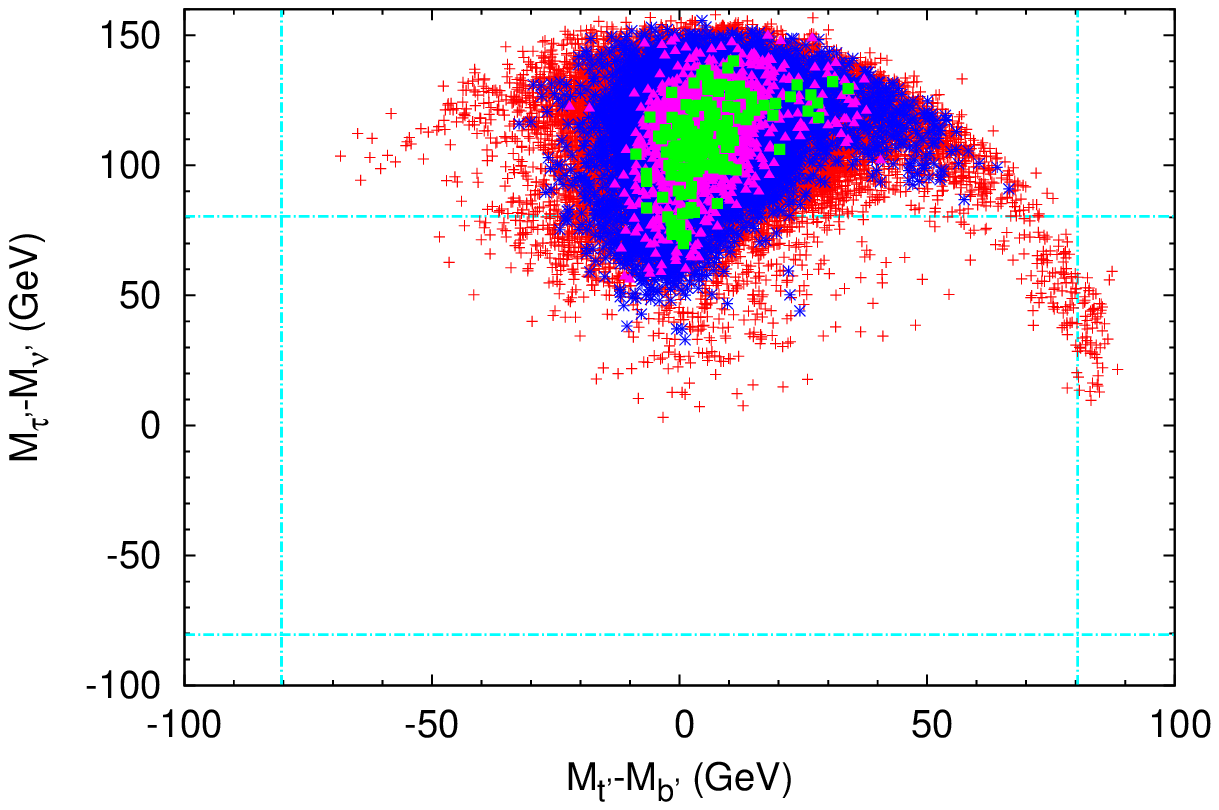}}
  \end{center}
  \caption{$M_{t'}$ v.s. $M_{b'}$ (upper left),
  $M_{\tau'}$ v.s. $M_{\nu'}$ (upper right).
  $M_{t'}$ v.s. $M_{\tau'}$ (lower left) and
  $M_{t'}-M_{b'}$ v.s. $M_{\tau'}-M_{\nu'}$ (lower right).
  We took $\sin(\beta-\alpha)=1$.
  The data points are the same as those in Fig.~\ref{mtp-mh-2HDM}.
  The blue lines in the upper figures correspond to 
  $M_{b'(\nu')}=M_{t'(\tau')}\pm M_W$.
  The blue lines in the lower right correspond to $\pm M_W$.
  \label{mtp-mbp-2HDM}}
\end{figure*}

\begin{figure}[t]
  \begin{center}
  \includegraphics{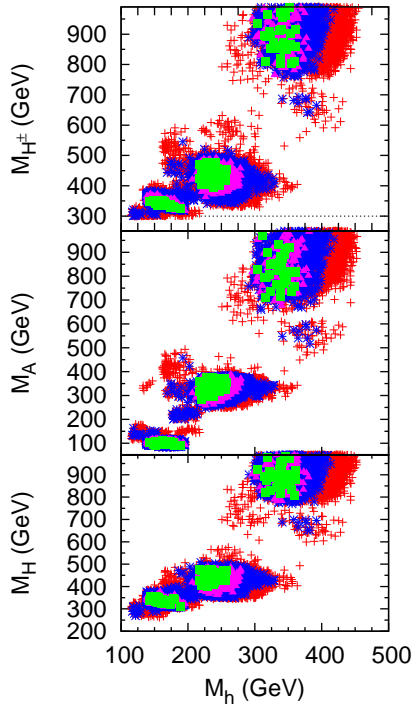}
  \end{center}
  \caption{Masses of the extra Higgs bosons.
  We took $\sin(\beta-\alpha)=1$.
  The data points are the same as those in Fig.~\ref{mtp-mh-2HDM}.
  \label{Higgs-mass-2HDM}}
\end{figure}

\begin{figure*}[t]
  \begin{center}
  \resizebox{0.4\textheight}{!}{\includegraphics{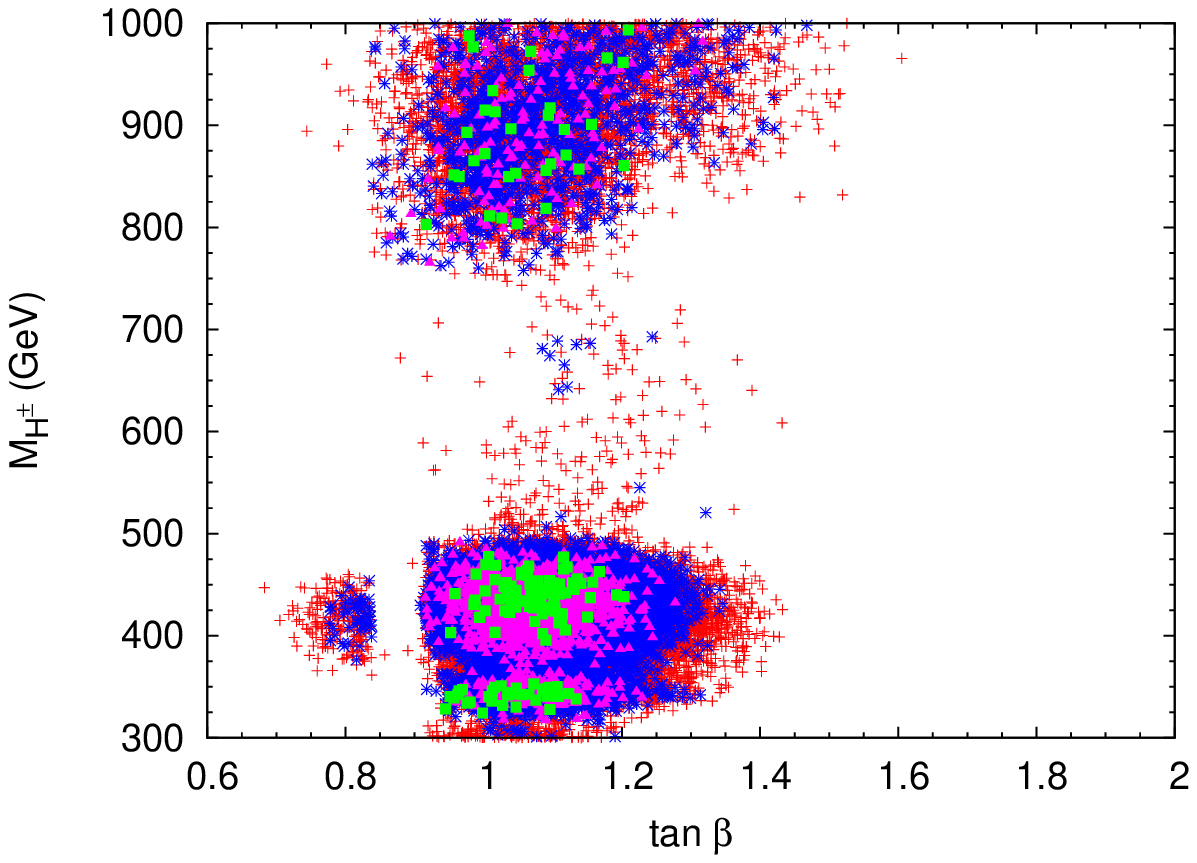}} \quad
  \resizebox{0.4\textheight}{!}{\includegraphics{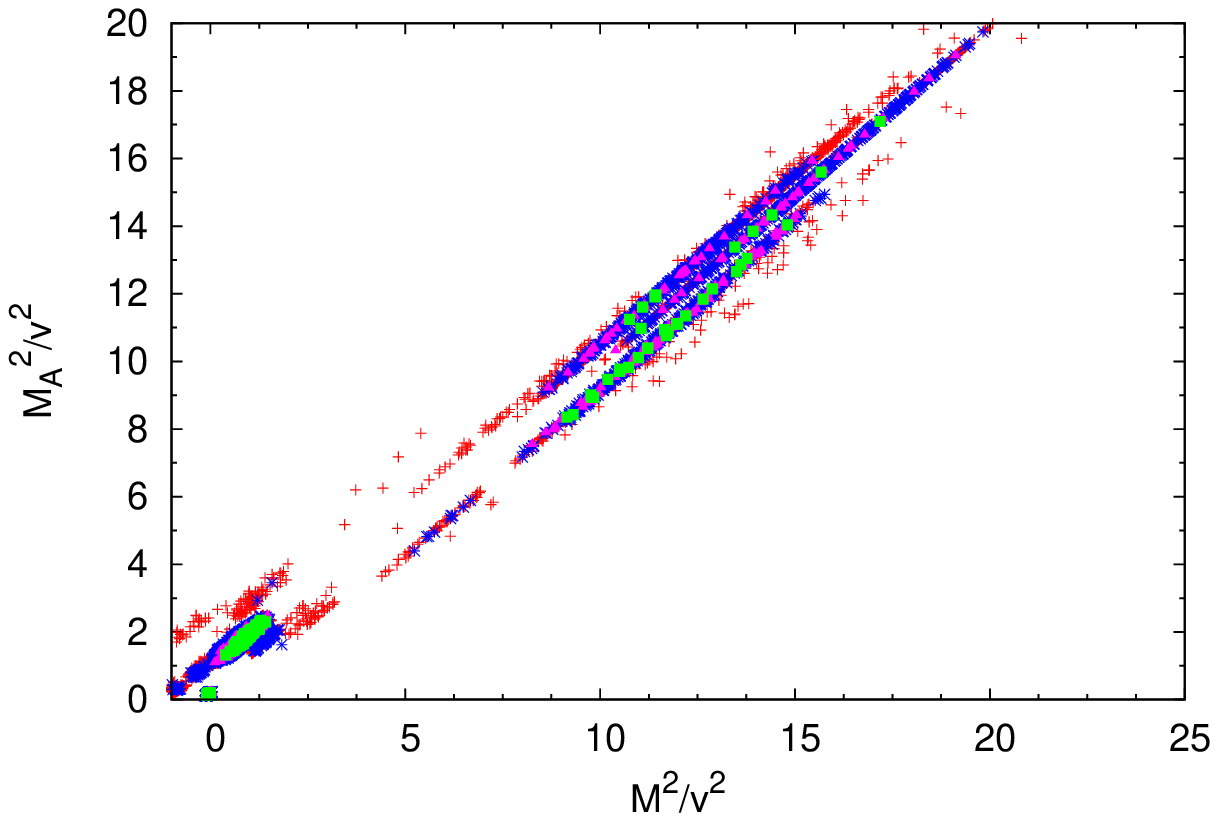}}
  \end{center}
  \caption{$\tan\beta$ v.s. $M_{H^\pm}$ and $M^2/v^2$ v.s. $M_A^2/v^2$.
  We took $\sin(\beta-\alpha)=1$.
  The data points are the same as those in Fig.~\ref{mtp-mh-2HDM}.
  We can read the value of $\lambda_5$ from the right figure
  by using Eq.~(\ref{lam-mass}).
  \label{H-tb-2HDM}}
\end{figure*}

\subsection{Numerical analysis}

The numerical analysis is similar to 
that in the previous section~\cite{Nie:1998yn}.

We vary the fermion masses, 
$256 \mbox{ GeV} < M_{t'} < 552 \mbox{ GeV}$,
$255 \mbox{ GeV} < M_{b'} < 552 \mbox{ GeV}$,
$100.8 \mbox{ GeV} < M_{\tau'} < 1.23 \mbox{ TeV}$ and
$90.3 \mbox{ GeV} < M_{\nu'} < 1.23 \mbox{ TeV}$.

For the Higgs sector, 
we vary the Higgs masses, $\tan\beta$ and $\lambda_5$.

The Higgs quartic couplings are theoretically constrained by 
the stability conditions for 
the Higgs potential~\cite{Deshpande:1977rw}
\footnote{
The factors of the quartic couplings $\lambda_1$ and $\lambda_2$
are different from those in 
Refs.~\cite{Deshpande:1977rw,Hill:1985tg,Kanemura:2004mg}.
},
\begin{subequations}
\begin{eqnarray}
&&  \lambda_1 > 0, \quad \lambda_2 > 0, \quad
    \lambda_3 > -2\sqrt{\lambda_1 \lambda_2}, \\
&&  \lambda_3 + \lambda_4 - |\lambda_5| > -2\sqrt{\lambda_1 \lambda_2}\,. 
\end{eqnarray}
\end{subequations}
and also the tree level unitarity 
bounds~\cite{Kanemura:1993hm,Kanemura:2004mg},
\begin{equation}
  |\tilde{a}_\pm|, |\tilde{b}_\pm|,|\tilde{c}_\pm|, |\tilde{d}_\pm|,
  |\tilde{e}_1|, |\tilde{e}_2|, |\tilde{f}_\pm|,
  |\tilde{f}_1|, |\tilde{f}_2| < 16 \pi \xi,
\end{equation}
with
\begin{subequations}
\begin{align}
&  \tilde{a}_\pm = 
   3(\lambda_1+\lambda_2)
   \pm \sqrt{9(\lambda_1-\lambda_2)^2+(2\lambda_3+\lambda_4)^2},\\
&  \tilde{b}_\pm = 
   (\lambda_1+\lambda_2)
   \pm \sqrt{(\lambda_1-\lambda_2)^2+\lambda_4^2},\\
&  \tilde{c}_\pm = \tilde{d}_\pm = 
    (\lambda_1+\lambda_2)
   \pm \sqrt{(\lambda_1-\lambda_2)^2+\lambda_5^2}, \\
&  \tilde{e}_1 = \lambda_3+2\lambda_4-3\lambda_5, \\
&  \tilde{e}_2 = \lambda_3-\lambda_5, \\
&  \tilde{f}_+ = \lambda_3+2\lambda_4+3\lambda_5, \\
&  \tilde{f}_- = \lambda_3+\lambda_5, \\
&  \tilde{f}_1 = \tilde{f}_2 = \lambda_3+\lambda_4 ,  
\end{align}
\end{subequations}
where we take $\xi = 1/2$, 
which corresponds to the radius of the Argand diagram.

Although we will ignore the mixing terms between the fourth generation
and the others, the charged Higgs mass should be severely constrained
by $b \to s \gamma$ and $R_b$, 
as in the three generation model~\cite{Cheung:2003pw}. 
In this paper, we do not fully analyze the experimental constraints.
Instead, we take $M_{H^\pm} \gtrsim 300$~GeV in order to evade 
the constraint from $b \to s \gamma$~\cite{Misiak:2006zs,Deschamps:2009rh}.
We also take into account 
the $R_b$-constraint for $\tan\beta < 1$~\cite{Denner:1991ie,Haber:1999zh}.
The $B^0$-$\bar{B}^0$ mixing yields less severer constraints only.
We do not consider too small or too large $\tan\beta$, 
because in such a case, the yukawa couplings reach so quickly 
the Landau pole.

Eventually, the parameter space for the Higgs sector
is taken as following;
$M_h^{\rm min} < M_{h} < 1 \mbox{ TeV}$,
$M_h < M_{H} < 1.5 \mbox{ TeV}$,
$300 \mbox{ GeV} < M_{H^\pm} < 1 \mbox{ TeV}$,
$93 \mbox{ GeV} < M_{A} < 1 \mbox{ TeV}$, 
$0.5 < \tan\beta < 5$, and $|\lambda_5| < \pi$,
where $M_h^{\rm min}$ corresponds to the lower bounds of $M_h$
for the various values of $\sin^2(\beta-\alpha)$ 
which can be read from the constraints of 
the LEP experiments~\cite{Barate:2003sz}.
In the case of $\sin(\beta-\alpha)=1$, 
it corresponds to the SM bound, $M_h^{\rm min}=114$~GeV.

Taking into account the RGE's for the yukawa and 
Higgs quartic couplings~\cite{Hill:1985tg,Komatsu:1981xh},
which are shown in Appendix~\ref{sec-RGE-2HDM}, 
and also imposing the instability bounds for the Higgs potential 
and the perturbative unitarity bounds on 
the yukawa and Higgs quartic couplings,
we calculate the cutoff $\Lambda$ at which some new physics or 
strong dynamics enters. 
The masses of the fermions and the Higgs bosons are 
also constrained by the $(S,T)$-parameters~\cite{He:2001tp}.

Since the parameter space is enormous,
we need to find efficiently the cutoff $\Lambda$ 
unlike in the analysis of the SM4. 
After generating the primary data with an equal probability, 
we refine all parameters so as to make $\Lambda$ larger. 
We cannot deny the possibility that 
we may overlook some favorable parameter region,
if the primary data might be too rough.

We depict the results in Figs.~\ref{mtp-mh-2HDM}--\ref{H-tb-2HDM}.

The relation between $M_{t'(\nu')}$ and $M_h$ is shown in 
Fig.~\ref{mtp-mh-2HDM}.
All data points are within 95\% C.L. limit of the $(S,T)$-parameters.
(See Fig.~\ref{S-T-2HDM}.)
Similarly, the mass relations between the fourth generation fermions 
are depicted in Fig.~\ref{mtp-mbp-2HDM}.
The masses of the extra heavy Higgs bosons are described in 
Fig.~\ref{Higgs-mass-2HDM}.
The allowed parameter region for $M_{H^\pm}$ and $\tan\beta$
is shown in the left of Fig.~\ref{H-tb-2HDM}.
The values of $\lambda_5$ can be read from the right of
Fig.~\ref{H-tb-2HDM} by using Eq.~(\ref{lam-mass}).

It is noticeable that the decay channels $h \to \bar{\nu}' \nu'$ 
and $\tau' \to \nu' + W^-$ are allowed in a wide parameter range.
(See the right of Fig.~\ref{mtp-mh-2HDM} and
the lower right of Fig.~\ref{mtp-mbp-2HDM}.)
Notice that $\tan\beta \simeq 1$ in order to relax 
the appearance of the Landau pole for the yukawa couplings.
(See the left of Fig.~\ref{H-tb-2HDM}.)

Schematically speaking, 
as shown in the left of Fig.~\ref{mtp-mh-2HDM} and 
Fig.~\ref{Higgs-mass-2HDM},
the results consist of high and low $M_h$ regimes,
$M_h \gtrsim M_{t'}$ and $M_h \lesssim M_{t'}$, respectively,
where we took the cutoff $\Lambda \gtrsim 2$~TeV.

We can confirm that the decoupling regime of the extra Higgs bosons, say,
$M_H \sim M_{H^\pm} \gtrsim 800$~GeV and $M_A \gtrsim 700$~GeV,
is contained in the region $M_h \gtrsim M_{t'}$.
(See Fig.~\ref{Higgs-mass-2HDM} and the right of
Fig.~\ref{H-tb-2HDM}.)
This is consistent with the analysis of the SM4.
In this case, the extra heavy Higgs bosons can decay into  
a quark/lepton pair of the fourth generation, 
if kinematically allowed.

A new feature of the two Higgs extension is thus characterized by 
the low $M_h$ regime, $M_h \lesssim M_{t'}$.
In the $(S,T)$-analysis, this regime is more favorable
than the high $M_h$ one, i.e., most of the data points 
inside the 68\% C.L. limit of the $(S,T)$-constraints
are for the former.
The point is that even for the low $M_h$, 
the Higgs potential can be stable 
owing to the dynamics of the Higgs quartic couplings.
When we take into account the Tevatron bounds of
the fourth generation quark masses, 
$M_{t'} > 311$~GeV and $M_{b'} > 338$~GeV~\cite{CDF-bound},
only a small parameter space is left, however.
Nevertheless, we here mention that 
the parameter region with $M_h \approx \mbox{100--300 GeV}$
is interesting, because the extra Higgs masses can be almost degenerate,
$M_{H} \sim M_{H^\pm} \sim M_A \sim \mbox{300--400 GeV}$,
and also a scenario with $M^2 = 0$ is possible
(see the right of Fig.~\ref{H-tb-2HDM}).
We also note that even in this regime, 
the leptonic decays of the Higgs bosons such as 
$h \to \bar{\nu}'\nu'$, $H^- \to \bar{\nu}' \tau'$ and etc. are open
in a certain parameter space.

We have analyzed only the case of $\sin(\beta-\alpha)=1$.
If we extend our analysis with a general $\sin(\beta-\alpha)$,
more favorable and exotic Higgs mass spectra can be found.
This will be performed elsewhere.

\section{Summary and discussions}
\label{summary}

We have reanalyzed the constraints on the mass spectrum of 
the fourth generation fermions and the Higgs bosons
for the SM4 and the THDM~II with $\sin(\beta-\alpha)=1$.
We showed that there are the noticeable correlations 
among the mass spectrum of the fermions and the Higgs bosons.

For the SM4, the favorable mass range of 
the physical Higgs boson $\phi^0$ is $M_{\phi^0} \gtrsim M_{t'}$
($M_{\phi^0} \gtrsim M_{t'} - \mbox{50 GeV}$)
for the cutoff $\Lambda \geq 2$~TeV ($\Lambda \geq 1$~TeV). 
We also found that the favorable parameter space is 
mainly contained in the region,
\begin{eqnarray}
\lefteqn{\hspace*{-1.5cm}
   \frac{3(M_{t'} - M_{b'})^2}{M_W^2}
 + \frac{(M_{\tau'} - M_{\nu'})^2}{M_W^2}
} \nonumber \\
&&
 \approx \mbox{(1.3--2.0)} + 1.4 \ln \frac{M_{\phi^0}}{M_{\phi^0,ref}} ,
\end{eqnarray}
and
\begin{equation}
  M_{t'} > M_{b'} , \qquad M_{\tau'} > M_{\nu'} \, . 
\end{equation}
(See also the semicircles in lower right of Fig.~\ref{mtp-mbp}.)
We showed the data samples corresponding to several scenarios
in Table~\ref{tab-1}.

For the THDM~II with $\sin(\beta-\alpha)=1$,
schematically speaking, 
there are two domains for the favorable mass range of 
the light CP even Higgs $h$, 
$M_h \gtrsim M_{t'}$ and $M_h \lesssim M_{t'}$.
The extra heavy Higgs decoupling regime is contained in the former.
This is consistent with the analysis of the SM4.
On the other hand, an almost degenerate Higgs mass spectrum 
such as $M_h \approx \mbox{100--300 GeV}$ and 
$M_{H} \sim M_{H^\pm} \sim M_A \sim \mbox{300--400 GeV}$
is allowed in the latter and in a part of the former. 
In this case, a model with $M^2=0$ is not excluded.
(See the right of Fig.~\ref{H-tb-2HDM}.)
As for the value of $\tan\beta$, we found that 
$\tan\beta \approx 1$ is favorable in the both domains. 

Concerning the decay channels of the charged leptons,
we found that $\tau' \to \nu' + W^-$ is allowed
in a wide parameter space in the SM4 and the THDM~II.
(See the lower right of Figs.~\ref{mtp-mbp} and \ref{mtp-mbp-2HDM}.)
The fourth generation quark masses can be degenerate $M_{t'}=M_{b'}$ 
or the decay channel $t' \to b' + W^{(*)}$ can be open, 
depending on the mass difference $M_{\tau'}-M_{\nu'}$.
(See the lower right of Figs.~\ref{mtp-mbp} and \ref{mtp-mbp-2HDM}.)
The Higgs $\phi^0$ in the SM4 and the light CP even Higgs $h$ in 
the THDM~II can decay into a pair of the fourth generation neutrinos.
(See the right of Figs.~\ref{mtp-mh} and \ref{mtp-mh-2HDM}.)
Furthermore, in the THDM~II,
a scenario that all Higgs bosons decay into 
a pair of the fourth generation leptons is possible, i.e.,
$h \to \bar{\nu}' \nu'$ and $H^- \to \bar{\nu}' \tau'$, etc..
(For studies of collider signals of the fourth generation leptons,
see, e.g., Ref.~\cite{CuhadarDonszelmann:2008jp}.)
We also comment that 
the main decay channel of the heavy CP even Higgs $H$
can be $H \to \bar{t}' t', \bar{b}' b'$,
if kinematically allowed~\cite{BarShalom:2010bh}.
Thus the phenomenology of the fourth generation models is very rich.

The implications of the analysis in this paper are obvious:
If the Tevatron and/or LHC discover $t'$ and/or $b'$,
the Higgs mass spectrum will be suggested, depending on the models.
On the other hand, if the LHC excludes
the $t'$ and/or $b'$ masses $M_{t',b'} \lesssim 500$~GeV at early stage,
a big parameter space will be gone.
In this case, essentially, a nonperturbative regime will be left
to be examined.

Many issues remain to be explored:

\begin{itemize}
\item We did not consider a general case of $\sin(\beta-\alpha)$.
There probably exist more favorable and exotic parameter regions
in the THDM~II. 
Moreover, we may consider a different yukawa structure other than 
the type~II~\cite{Barger:1989fj}. 

\item Majorana neutrinos can reduce $S$ and $T$~\cite{Bertolini:1990ek}.
It may affect the mass spectrum of the fourth generation fermions
and the Higgs bosons.

\item The two-loop effects are probably relevant 
for more precise predictions of the mass spectrum.
The theoretical lower bound for the Higgs mass,
which essentially corresponds to the instability bound 
of the Higgs potential, will be almost unchanged, however,
because the parameters certainly stay in a perturbative region.

\item 
We did not take into account the mixing angle between 
the fourth and third generations.
This is, of course, very important to discuss 
the flavor constraints and realistic decay chains of 
the fourth generation quarks and 
leptons~\cite{Bobrowski:2009ng,Chanowitz:2009mz}.

\item If the main branching ratios of $t'$ and $b'$ are 
different from well-studied ones in experiments,
a first evidence of the fourth generation might be found
in the Higgs physics, for example, as a huge enhancement
of the golden mode, $gg \to \phi^0/h \to ZZ$.
Concerning the loop induced processes,
it is important to notice that 
the loop effects in $h \to gg$, $h \to \gamma\gamma$ and
$A \to gg, \gamma\gamma$ are quite 
different~\cite{Higgs-hunter}.
Related to such possibilities,
there should exist very large non-decoupling effects 
in the triple Higgs coupling arising from the fourth generation 
quarks and leptons~\cite{Kanemura:2002vm,Kanemura:2004mg}.
The triple Higgs coupling is testable at 
the LHC/vLHC/sLHC~\cite{Baur:2002rb} and at the ILC~\cite{Ilyin:1995iy}.

\item The fourth generation model may play an important role
in B-CP asymmetries~\cite{Hou:2006zza,Soni:2008bc}
and also in the electroweak baryogenesis~\cite{Carena:2004ha}.

\item Last but not least, 
if only a nonperturbative regime is left in the near future,
what kind of study will be relevant?
For example, when the fourth generation quarks are ultraheavy
beyond the perturbative unitarity bound,
is there some nonperturbative effect
in the nonresonant $gg \to ZZ$ process~\cite{Chanowitz:1992rf}?
\end{itemize}

\acknowledgments

The author thanks to S. Kanemura for fruitful discussions
and to V.A.Miransky for useful comments.
This work is supported by
the Grant-in-Aid for Science Research, Ministry of Education, 
Culture, Sports, Science and Technology, Japan, No. 16081211.

\appendix

\section{RGE's for the SM4}
\label{RGE-sm4}

For the gauge couplings, the RGE's are 
\begin{equation}
(16 \pi^2) \mu \frac{\partial }{\partial \mu} g_i = - c_i g_i^3,  
\end{equation}
with
\begin{eqnarray}
c_1 &=& -\frac{1}{6}N_H-\frac{20}{9}N_g, \\
c_2 &=& \frac{22}{3}-\frac{4}{3}N_g-\frac{1}{6} N_H, \\
c_3 &=& 11 - \frac{4}{3}N_g,
\end{eqnarray} 
where $N_g$ and $N_H$ denote the number of generations and
the number of Higgs doublets, respectively. 
Although we did not show explicitly the formulae, 
we take into account the threshold effects.

The RGE's of the yukawa couplings are~\cite{Machacek:1983tz,Hill:1980sq}
\begin{subequations}
\begin{eqnarray}
(16 \pi^2) \mu \frac{\partial }{\partial \mu} y_{t} &=&
 - (8g_3^2 + \frac{9}{4}g_2^2+\frac{17}{12}g_1^2)y_t
 + \frac{9}{2}y_t^3    \nonumber \\ && \hspace*{-2.3cm}
 + y_t \bigg[\,3 y_{t'}^2 + 3 y_{b'}^2 + \frac{3}{2} y_b^2
    + y_{\nu'}^2 + y_{\tau'}^2 + y_{\tau}^2\,\bigg], 
\end{eqnarray}
\begin{eqnarray}
(16 \pi^2) \mu \frac{\partial }{\partial \mu} y_{b} &=&
 - (8g_3^2 + \frac{9}{4}g_2^2+\frac{5}{12}g_1^2)y_{b}
 + \frac{9}{2}y_{b}^3    \nonumber \\ && \hspace*{-2.3cm}
 + y_{b} \bigg[\,3 y_{b'}^2 + 3 y_{t'}^2 + \frac{3}{2}y_{t}^2
    + y_{\nu'}^2 + y_{\tau'}^2 + y_{\tau}^2\,\bigg], 
\end{eqnarray}
\begin{eqnarray}
(16 \pi^2) \mu \frac{\partial }{\partial \mu} y_{\tau} &=&
 - (\frac{9}{4}g_2^2+\frac{15}{4}g_1^2)y_{\tau}
 + \frac{5}{2}y_{\tau}^3    \nonumber \\ && \hspace*{-2.3cm}
 + y_{\tau} \bigg[\,3 y_{t'}^2 + 3 y_{b'}^2 + 3 y_{t}^2 + 3 y_{b}^2
 + y_{\nu'}^2 + y_{\tau'}^2\,\bigg], 
\end{eqnarray}
\begin{eqnarray}
(16 \pi^2) \mu \frac{\partial }{\partial \mu} y_{t'} &=&
 - (8g_3^2 + \frac{9}{4}g_2^2+\frac{17}{12}g_1^2)y_{t'}
 + \frac{9}{2}y_{t'}^3    \nonumber \\ && \hspace*{-2.3cm}
 + y_{t'} \bigg[\,\frac{3}{2} y_{b'}^2 + 3 y_{t}^2
    + 3 y_b^2 + y_{\nu'}^2 + y_{\tau'}^2 + y_{\tau}^2\,\bigg], 
\end{eqnarray}
\begin{eqnarray}
(16 \pi^2) \mu \frac{\partial }{\partial \mu} y_{b'} &=&
 - (8g_3^2 + \frac{9}{4}g_2^2+\frac{5}{12}g_1^2)y_{b'}
 + \frac{9}{2}y_{b'}^3   \nonumber \\ && \hspace*{-2.3cm}
 + y_{b'} \bigg[\,\frac{3}{2}y_{t'}^2 + 3 y_{t}^2 + 3 y_{b}^2 
    + y_{\nu'}^2 + y_{\tau'}^2 + y_{\tau}^2\,\bigg], 
\end{eqnarray}
\begin{eqnarray}
(16 \pi^2) \mu \frac{\partial }{\partial \mu} y_{\nu'} &=&
 - (\frac{9}{4}g_2^2+\frac{3}{4}g_1^2)y_{\nu'}
 + \frac{5}{2}y_{\nu'}^3   \nonumber \\ && \hspace*{-2.5cm}
 + y_{\nu'} \bigg[\,3 y_{t'}^2 + 3 y_{b'}^2 + 3 y_{t}^2 + 3 y_{b}^2
 - \frac{1}{2}y_{\tau'}^2 + y_{\tau}^2\,\bigg], 
\end{eqnarray}
\begin{eqnarray}
(16 \pi^2) \mu \frac{\partial }{\partial \mu} y_{\tau'} &=&
 - (\frac{9}{4}g_2^2+\frac{15}{4}g_1^2)y_{\tau'}
 + \frac{5}{2}y_{\tau'}^3   \nonumber \\ && \hspace*{-2.5cm}
 +  y_{\tau'} \bigg[\,3 y_{t'}^2 + 3 y_{b'}^2 + 3 y_{t}^2 + 3 y_{b}^2
 - \frac{1}{2}y_{\nu'}^2 + y_{\tau}^2\,\bigg], 
\end{eqnarray}
\end{subequations}
where we did not show explicitly the threshold effects of the fermions.
Inside of the square bracket, $y_f^2$ $(f=t',b',\cdots)$ should be 
regarded as $y_f^2 \theta (\mu - m_f)$, where $m_f$ is 
the corresponding $\overline{\rm MS}$-mass of the fermions.

The RGE for the Higgs quartic coupling is given by
\begin{eqnarray}
&&
  (16 \pi^2) \mu \frac{\partial }{\partial \mu} \lambda =
   24 \lambda^2  - 3 \lambda (3 g_2^2 + g_1^2) \nonumber \\
&& \qquad 
   + 4\lambda \bigg[\,3(y_{t'}^2 + y_{b'}^2 + y_t^2 + y_b^2)
   + y_{\tau'}^2 + y_{\nu'}^2 + y_{\tau}^2\,\bigg] \nonumber \\
&& \qquad
 - 2 \bigg[\,3(y_{t'}^4 + y_{b'}^4 + y_t^4 + y_b^4)
    + y_{\tau'}^4 + y_{\nu'}^4 + y_{\tau}^4\,\bigg] \nonumber \\
&& \qquad
    + \frac{3}{8}\bigg[\,2 g_2^4 +(g_2^2 + g_1^2)^2\,\bigg]\, .
\end{eqnarray}

\section{RGE's for the THDM~II}
\label{sec-RGE-2HDM}

Let us consider the RGE's for the THDM~II.

The RGE's for Yukawa couplings are given by~\cite{Hill:1985tg,Komatsu:1981xh}
\begin{subequations}
\begin{eqnarray}
&&
   (16 \pi^2) \mu \frac{\partial }{\partial \mu} y_{t} =
 - (8g_3^2 + \frac{9}{4}g_2^2+\frac{17}{12}g_1^2)y_t
 + \frac{9}{2}y_t^3 \nonumber \\
&& \qquad \qquad \qquad
 + y_t \bigg[\,3 y_{t'}^2 + \frac{1}{2}y_b^2
    + y_{\nu'}^2 \,\bigg], 
\end{eqnarray}
\begin{eqnarray}
&&
(16 \pi^2) \mu \frac{\partial }{\partial \mu} y_{b} =
 - (8g_3^2 + \frac{9}{4}g_2^2+\frac{5}{12}g_1^2)y_{b}
 + \frac{9}{2}y_{b}^3 \nonumber \\
&&  \qquad \qquad \qquad
 + y_{b} \bigg[\,3 y_{b'}^2 + \frac{1}{2}y_{t}^2
    + y_{\tau'}^2 + y_{\tau}^2 \,\bigg], 
\end{eqnarray}
\begin{eqnarray}
&&
(16 \pi^2) \mu \frac{\partial }{\partial \mu} y_{\tau} =
 - (\frac{9}{4}g_2^2+\frac{15}{4}g_1^2)y_{\tau}
 + \frac{5}{2}y_{\tau}^3 \nonumber \\
&&  \qquad \qquad \qquad
 + y_{\tau} \bigg[\,3 y_{b}^2 + 3 y_{b'}^2 + y_{\tau'}^2\,\bigg], 
\end{eqnarray}
\begin{eqnarray}
&&
(16 \pi^2) \mu \frac{\partial }{\partial \mu} y_{t'} =
 - (8g_3^2 + \frac{9}{4}g_2^2+\frac{17}{12}g_1^2)y_{t'}
 + \frac{9}{2}y_{t'}^3 \nonumber \\
&&  \qquad \qquad \qquad
 + y_{t'} \bigg[\,3 y_{t}^2 + \frac{1}{2}y_{b'}^2
    + y_{\nu'}^2 \,\bigg], 
\end{eqnarray}
\begin{eqnarray}
&&
(16 \pi^2) \mu \frac{\partial }{\partial \mu} y_{b'} =
 - (8g_3^2 + \frac{9}{4}g_2^2+\frac{5}{12}g_1^2)y_{b'}
 + \frac{9}{2}y_{b'}^3 \nonumber \\
&&  \qquad \qquad \qquad
 + y_{b'} \bigg[\,3 y_{b}^2 + \frac{1}{2}y_{t'}^2
    + y_{\tau'}^2 + y_{\tau}^2\,\bigg], 
\end{eqnarray}
\begin{eqnarray}
&&
(16 \pi^2) \mu \frac{\partial }{\partial \mu} y_{\nu'} =
 - (\frac{9}{4}g_2^2+\frac{3}{4}g_1^2)y_{\nu'}
 + \frac{5}{2}y_{\nu'}^3 \nonumber \\
&&  \qquad \qquad
 + y_{\nu'} \bigg[\,3 y_{t}^2 + 3 y_{t'}^2 + \frac{1}{2}y_{\tau'}^2\,\bigg], 
\end{eqnarray}
\begin{eqnarray}
&&
(16 \pi^2) \mu \frac{\partial }{\partial \mu} y_{\tau'} =
 - (\frac{9}{4}g_2^2+\frac{15}{4}g_1^2)y_{\tau'}
 + \frac{5}{2}y_{\tau'}^3 \nonumber \\
&&  \qquad \qquad
 + y_{\tau'} \bigg[\,3 y_{b}^2 + 3 y_{b'}^2 + \frac{1}{2}y_{\nu'}^2
                    + y_{\tau}^2\,\bigg], 
\end{eqnarray}
\end{subequations}
where we ignored $y_{c}$, $y_{\nu}$, etc..

The RGE's for the Higgs quartic self-couplings 
are~\cite{Hill:1985tg,Komatsu:1981xh}
\begin{eqnarray}
&&
  (16 \pi^2) \mu \frac{\partial }{\partial \mu} \lambda_1 =
   24\lambda_1^2+2\lambda_3^2+2\lambda_3\lambda_4+\lambda_4^2
  +\lambda_5^2 \nonumber \\
&&  \qquad \qquad
  -3\lambda_1(3g_2^2+g_1^2)
  +\frac{3}{8}\bigg[\,2 g_2^4 + (g_2^2+g_1^2)^2\,\bigg]  \nonumber \\
&&  \qquad \qquad
 +4\lambda_1\bigg[\, 3 y_{b'}^2 + 3 y_{b}^2 + y_{\tau'}^2
                     + y_{\tau}^2\,\bigg]\nonumber \\
&&  \qquad \qquad
   -2 \bigg[\,3 y_{b'}^4 + 3 y_{b}^4 + y_{\tau'}^4
              + y_{\tau}^4\,\bigg], 
\end{eqnarray}
\begin{eqnarray}
&&
   (16 \pi^2) \mu \frac{\partial }{\partial \mu} \lambda_2 = 
    24\lambda_2^2+2\lambda_3^2+2\lambda_3\lambda_4+\lambda_4^2
  +\lambda_5^2\nonumber \\
&&   \qquad \qquad
  -3\lambda_2(3g_2^2+g_1^2)
  +\frac{3}{8}\bigg[\,2 g_2^4 + (g_2^2+g_1^2)^2\,\bigg]  \nonumber \\
&&  \qquad \qquad
  +4\lambda_2\bigg[\, 3 y_{t'}^2 + 3 y_{t}^2 + y_{\nu'}^2\,\bigg]\nonumber \\
&&  \qquad \qquad
   -2 \bigg[\,3 y_{t'}^4 + 3 y_{t}^4 + y_{\nu'}^4\,\bigg], 
\end{eqnarray}
\begin{eqnarray}
&&
   (16 \pi^2) \mu \frac{\partial }{\partial \mu} \lambda_3 = 
   2(\lambda_1+\lambda_2)(6\lambda_3+2\lambda_4)+4\lambda_3^2
  +2\lambda_4^2 \nonumber \\
&&  \qquad 
  +2\lambda_5^2 -3\lambda_3(3g_2^2+g_1^2)
  +\frac{3}{4}\bigg[\,2 g_2^4 + (g_2^2-g_1^2)^2\,\bigg]  \nonumber \\
&&  \qquad 
  +2\lambda_3\bigg[\, 3 (y_{t}^2 + y_{b}^2 + y_{t'}^2 + y_{b'}^2)
                 + y_{\nu'}^2 +  y_{\tau'}^2 + y_{\tau}^2\,\bigg]\nonumber \\
&&  \qquad 
  -4 \bigg[\,3 y_{t'}^2 y_{b'}^2 + 3 y_{t}^2 y_{b}^2 + y_{\nu'}^2 y_{\tau'}^2
      \,\bigg], 
\end{eqnarray}

\begin{eqnarray}
&&
   (16 \pi^2) \mu \frac{\partial }{\partial \mu} \lambda_4 = 
   4(\lambda_1+\lambda_2 + 2\lambda_3+\lambda_4)\lambda_4+8\lambda_5^2
    \nonumber \\
&&  \qquad \qquad
  -3\lambda_4(3g_2^2+g_1^2) + 3g_1^2g_2^2  \nonumber \\
&&  \qquad \qquad
  +2\lambda_4\bigg[\, 3 (y_{t}^2 + y_{b}^2 + y_{t'}^2 + y_{b'}^2)
       + y_{\nu'}^2 +  y_{\tau'}^2 + y_{\tau}^2\,\bigg]\nonumber \\
&&  \qquad \qquad
  +4 \bigg[\,3 y_{t'}^2 y_{b'}^2 + 3 y_{t}^2 y_{b}^2 + y_{\nu'}^2 y_{\tau'}^2
      \,\bigg], 
\end{eqnarray}
\begin{eqnarray}
&&
    (16 \pi^2) \mu \frac{\partial }{\partial \mu} \lambda_5 = 
    \lambda_5\bigg[\,4(\lambda_1+\lambda_2)+8\lambda_3+12\lambda_4
   \nonumber \\
&&  \qquad \qquad
   -3(3g_2^2+g_1^2)
   +2\bigg\{\, 3 (y_{t}^2 + y_{b}^2 + y_{t'}^2 + y_{b'}^2)
   \nonumber \\
&&  \qquad \qquad
       + y_{\nu'}^2 +  y_{\tau'}^2 + y_{\tau}^2\,\bigg\}\,\bigg].
\end{eqnarray}
Note that the definitions for $\lambda_1$ and $\lambda_2$ are 
twice larger than those in Ref.~\cite{Hill:1985tg}.

\end{document}